%% file: main.tex
\documentclass[a4paper,UKenglish,cleveref,autoref,thm-restate]{lipics-v2021}

\hideLIPIcs

\usepackage[T1]{fontenc}
\usepackage{textcomp}
\usepackage{amsmath,amsthm}
\usepackage{xspace}

\usepackage{color,enumerate,array,graphicx,tabularx,listings}
\usepackage[table]{xcolor}

\usepackage{thm-restate}

\usepackage{algorithm,float,algorithmicx}
\usepackage[noend]{algpseudocode}

\usepackage{dsfont}
\usepackage{forest}

\usetikzlibrary{graphs}
\usetikzlibrary{graphs.standard}

\newcounter{SaveAlgLine}

\algnewcommand\algorithmiccase{\textbf{case}}

\algnewcommand\StateCase[1]{%
    \State\hphantom{x}\ #1 \algorithmicthen
}

\algdef{SE}[CASE]{Case}{EndCase}[1]
{\algorithmiccase\ #1\ \algorithmicthen}
{\algorithmicend\ \algorithmiccase}

\algdef{SE}{Begin}{End}
{\textbf{begin}}
{\textbf{end}}

\algnewcommand{\IfL}[1]{%
    \State\algorithmicif\ #1\ \algorithmicthen
}

\algnewcommand{\EndIfL}{%
    \unskip\ \algorithmicend\ \algorithmicif
}

\newif\ifannote

\annotetrue

\ifannote
    \newcommand{\anninsert}[2]{{\color{#1}#2}}
    \newcommand{\anncomment}[3]{{\color{#1}[#2: #3]}}
\else
    \newcommand{\anninsert}[2]{#2}
    \newcommand{\anncomment}[3]{}
\fi

\newcommand{\different}[1]{%
    \anninsert{blue}{#1}
}

\newtheorem{invariant}{Invariant}

\title{Asynchronous Byzantine Lattice Agreement
with Logarithmic Response Time and Optimal Resilience}

\titlerunning{Byzantine Lattice Agreement}

\author
{Hagit Attiya}
{Technion, Haifa, Israel}
{hagit@cs.technion.ac.il}
{https://orcid.org/0000-0002-8017-6457}
{Supported by the Israel Science Foundation (22/1425 and 25/1849).}

\author
{Yao David Kolegain}
{Aix-Marseille University, CNRS, LIS, Marseille, France}
{yao-david.kolegain@etu.univ-amu.fr}
{https://orcid.org/0009-0008-5703-4988}
{is supported by a French government grant managed by the Agence Nationale de la Recherche under the France 2030 program, reference ANR-23-PECL-0009.}

\author
{Arnaud Labourel}
{Aix-Marseille University, CNRS, LIS, Marseille, France}
{arnaud.labourel@lis-lab.fr}
{https://orcid.org/0000-0003-0162-1899}
{is supported by a French government grant managed by the Agence Nationale de la Recherche under the France 2030 program, reference ANR-23-PECL-0009.}

\author
{Alessia Milani}
{Aix-Marseille University, CNRS, LIS, Marseille, France}
{alessia.MILANI@univ-amu.fr}
{https://orcid.org/0009-0005-6459-6725}
{is supported by a French government grant managed by the Agence Nationale de la Recherche under the France 2030 program, reference ANR-23-PECL-0009.}

\authorrunning{H. Attiya, Y. D. Kolegain,
A. Labourel, and A. Milani}

\Copyright{Hagit Attiya, Yao David Kolegain,
Arnaud Labourel, and Alessia Milani}



\nolinenumbers

\begin{document}

\maketitle 

\begin{abstract}
Lattice agreement is a fundamental coordination primitive in which processes decide on comparable values rather than identical ones. 
In the asynchronous message-passing setting with Byzantine failures, existing algorithms either require authentication or do not simultaneously achieve optimal resilience and logarithmic response time.

We present an asynchronous Byzantine lattice agreement algorithm that tolerates up to $f$ Byzantine failures among $n>3f$ processes, requires no authentication, and completes in $O(\log f)$ time. 
The algorithm builds on the classifier-tree framework of Attiya and Rachman, replacing shared-memory communication with quorum-based message passing.

The main challenge is preserving the information-flow invariants underlying the classifier tree despite Byzantine behavior. To address this challenge, we introduce \emph{Supported Reliable Broadcast} (SRB), a communication abstraction that combines reliable broadcast with protocol-specific admissibility checks. 
Messages are disseminated using reliable broadcast, but are delivered to the protocol only after verifying that they are consistent with previously propagated information about the sender's state. 
This prevents Byzantine processes from bypassing the information-flow guarantees required by the classifier tree.

A central technical difficulty lies in breaking the apparent circular dependency between message delivery and admissibility: admissibility depends on local state established by previously delivered messages, while message delivery itself depends on admissibility. 
Our proof resolves this circularity by separating reliable dissemination from admissibility propagation, yielding a modular correctness argument. 
\end{abstract}

\input{introduction}

\input{related}
\input{preliminaries}

\input{SRB}
\input{one-shot-BLA}

\section{Summary}

We presented an asynchronous Byzantine lattice agreement algorithm that completes in $O(\log f)$ time;
it only assumes $n>3f$ and does not require authentication.  
The algorithm combines the classifier-tree framework of Attiya and Rachman with quorum-based message passing and a novel supported reliable broadcast mechanism. 
SRB enforces the information-flow constraints needed to preserve the classifier invariants despite Byzantine failures.
A natural open question is to improve the time bound or to prove that it is optimal.

Beyond the specific lattice-agreement construction, supported reliable broadcast provides a way to combine reliable dissemination with protocol-specific admissibility conditions.  
It may be useful in other Byzantine-tolerant algorithms in which messages must be justified by information accumulated along different execution paths.

\bibliography{references}

\end{document}

%% file: introduction.tex
\section{Introduction}

\emph{Lattice agreement}~\cite{Attiya_atomicHR1995} allows processes to propose arbitrary inputs and decide on output values that are \emph{comparable} in a lattice.
A typical case of lattice agreement considers the \emph{subsets lattice}, 
with the usual subset and inclusion relations. 
Lattice agreement is weaker than consensus, where the outputs must be identical, and it can be solved in settings where consensus is impossible or substantially more expensive. 
Beyond its intrinsic theoretical interest, lattice agreement is an important building block for distributed objects and monotone replicated computations, including atomic snapshots~\cite{Attiya_atomicHR1995}, CRDTs~\cite{ShapiroPBZ2011,zheng_linearizable_2018}, and related applications.

One of the central correctness requirements in lattice agreement is that the outputs are joins of the input values; in the special case of the subsets lattice, this means that the output sets contain only values proposed by processes. 
(This condition is called \emph{upward validity}.) 

Originally, lattice agreement was studied in the context of crash failures, either in shared memory~\cite{attiya_atomic_1998} or message passing~\cite{zheng_linearizable_2018}.
Due to its applicability for state machine replication~\cite{zheng_linearizable_2018}, 
lattice agreement was recently studied also in the presence of \emph{Byzantine} failures~\cite{zheng_byzantine_2020_asynchronous}. 
Byzantine failures require us to relax \emph{upward validity} so that $f$ faulty processes can introduce at most $f$ arbitrary, but fixed, values into the output sets. 

Several Byzantine lattice agreement algorithms were presented, 
in synchronous and asynchronous settings, with and without authentication 
(see Table~\ref{table:related work} and Section~\ref{sec:related}). 
However, no \emph{asynchronous} algorithm simultaneously achieves optimal resilience ($n > 3f$) 
and logarithmic response time, without using authentication.
The goal of this paper is to close this gap.

To understand why this goal is challenging, 
consider the lattice agreement algorithm of Attiya and Rachman~\cite{attiya_atomic_1998}; 
this is a wait-free shared-memory algorithm that works in $O(n \log n)$ operations. 
The algorithm is constructed as a \emph{tree of classifiers}.
Starting at the root classifier, processes are repeatedly partitioned into \emph{low-} and \emph{high-information} groups. 
Low processes retain their current information, whereas high processes obtain information that subsumes that of all low processes. 
The key invariant is that whenever processes split into low and high groups, every value known to a low process is eventually incorporated into the state of every high process.
The same classification procedure is then applied recursively within each subgroup. 
Consequently, different processes gradually follow different paths in the classifiers tree, 
maintaining different local states and executing different classifier instances.

In the original algorithm, once a process writes information to the shared memory, it is never removed and later readers can observe it. 
In a message-passing algorithm, a Byzantine process classified 
as low may later introduce information that is never observed 
by the corresponding high processes, 
violating a key invariant underlying the classifier tree. 
In general, a Byzantine process may claim to belong to a classifier or to have followed a particular path without executing the corresponding protocol.
This necessitates a way to ensure that every delivered message is consistent with the sender's position in the classifier tree.

We address this problem with \emph{Supported Reliable Broadcast} (SRB). 
Messages are first disseminated using standard Byzantine reliable broadcast~\cite{bracha_asynchronous_1987}. 
However, before a delivered message is passed to the protocol, it is checked against a protocol-specific admissibility predicate that verifies that the message is consistent with previously delivered information about the sender. 
This communication layer guarantees not only reliable dissemination, but also that every message delivered could have been generated by a correct process following the protocol.
Admissibility may appear to create a circular dependency: message delivery depends on admissibility, while admissibility depends on protocol state that is itself established by previously delivered messages. 
We resolve this by separating reliable dissemination from admissibility propagation, allowing the latter to be proved inductively over the classifier tree.

Support-based message filtering has appeared in earlier Byzantine protocols;
we discuss these mechanisms in Section~\ref{sec:related}.
The classifier tree makes this approach substantially more complicated:
different processes follow different execution paths, so admissibility must justify both the information carried by a message and the sender's claimed position in the tree.

Starting from the Attiya–Rachman classifier tree, we use standard quorum techniques originating with the message-passing simulation of shared memory~\cite{AttiyaBD1995}. 
Supported reliable broadcast enforces the information-flow guarantees required by the classifier tree. An initial dissemination phase ensures that every correct process begins the classifier tree with at least $n-f$ values, reducing the depth of the classifier tree from $O(\log n)$ to $O(\log f)$. The resulting algorithm achieves logarithmic response time while preserving the classifier invariants in the presence of Byzantine processes.
The proof cleanly separates communication guarantees, admissibility propagation, and classifier properties.

In summary, our contributions are three-fold:
\begin{itemize}
\item  
Algorithmically, 
we present the first asynchronous message-passing Byzantine lattice agreement algorithm that simultaneously achieves optimal resilience ($n>3f$), logarithmic response time ($O(\log f)$), and does not require authentication.

\item 
Conceptually, we formulate supported reliable broadcast as a communication abstraction separating reliable dissemination from protocol-specific admissibility checks, making it possible to support non-uniform classifier-based protocols.

\item 
Technically, we develop a proof that separates communication guarantees, admissibility propagation, and classifier properties.
\end{itemize}

%% file: related.tex
\section{Related Work}
\label{sec:related}

\begin{table}[bt]
\centering
\small
\begin{tabular}{|l|l|l|c|c|c|c|}
\hline
 &
 &
\textbf{Failures} &
\textbf{Authen.} &
\textbf{Resilience} &
\textbf{Time / Rounds} \\

\hline
\cite{Attiya_atomicHR1995}
& Synchronous
& Crash
& No
& $n > f$
& $O(\log n)$ \\

\hline

\cite{zheng_byzantine_2020_synchronous}
& Synchronous
& Byzantine
& No
& $n > 3f$
& $O(\log f)$ \\
\hline\hline

\cite{zheng_linearizable_2018}
& Asynchronous  
& Crash
& No
& $n > 2f$
& $O(\log f)$ \\
\hline

\cite{di_luna_byzantine_2020}
& Asynchronous
& Byzantine
& No
& $n > 3f$
& $O(f)$ \\
\hline

\cite{zheng_byzantine_2020_asynchronous}
& Asynchronous
& Byzantine
& No
& $n > 5f$
& $O(\log f)$ \\
\hline

\cite{zheng_byzantine_2020_asynchronous}
& Asynchronous
& Byzantine
& Yes
& $n > 3f$
& $O(\log f)$ \\
\hline

\rowcolor{gray!25}
\textbf{This paper}
& Asynchronous
& Byzantine
& No
& $n > 3f$
& $O(\log f)$ \\
\hline
\end{tabular}
\caption{Selected lattice agreement algorithms in message-passing systems.}\label{table:related work}
\end{table}

\subparagraph*{Crash-tolerant lattice agreement.}
Lattice agreement was introduced by Attiya, Herlihy, and
Rachman~\cite{Attiya_atomicHR1995}.
Their synchronous message-passing algorithm uses the basic idea of
recursively partitioning processes and directing information from one
part to the other, obtaining an $O(\log n)$ round complexity.
The partition is determined statically, according to process identities.
Attiya and Rachman~\cite{attiya_atomic_1998} developed a dynamic
version of this approach: processes are classified according to the
information they acquire during the execution.
The resulting classifier framework underlies their wait-free
shared-memory algorithm, with $O(n\log n)$ step complexity, and is
the basis for our algorithm.

For asynchronous message-passing systems, 
lattice agreement requires $n>2f$, by a standard partitioning argument.
Faleiro, Rajamani, Rajan, Ramalingam, and Vaswani~\cite{faleiro_generalized_2012}
gave the first lattice agreement algorithm for such systems, with linear time complexity.
Zheng, Garg, and Kaippallimalil~\cite{zheng_linearizable_2018}
later adapted the classifier approach to asynchronous message passing,
obtaining an $O(\log f)$-time algorithm for the optimal resilience $n>2f$.

\subparagraph*{Byzantine lattice agreement.}
Di Luna, Anceaume, and Querzoni~\cite{di_luna_byzantine_2020} formulated the 
Byzantine upward-validity condition used in this paper, 
and showed that $n>3f$ is necessary. 
Their asynchronous unauthenticated algorithm matches this resilience bound and has $O(f)$ time complexity.
For synchronous systems, Di Luna, Anceaume, Bonomi, and Querzoni~\cite{di_luna_synchronous_2020}
gave $O(\log f)$-time algorithms for $n>4f$ without authentication
and for $n>3f$ with authentication.
Zheng and Garg~\cite{zheng_byzantine_2020_synchronous} subsequently obtained an unauthenticated 
synchronous algorithm with $O(\log f)$ time complexity for the optimal resilience $n>3f$.

The $O(\log f)$-time algorithms for asynchronous systems, 
by Zheng and Garg~\cite{zheng_byzantine_2020_asynchronous}, 
either use authentication for the optimal resilience $n>3f$
or avoid authentication and assume $n>5f$.
Thus, without authentication, 
prior asynchronous algorithms achieved either optimal resilience or logarithmic time, 
but not both.
\emph{Our algorithm closes this gap,}
achieving $O(\log f)$ time for $n>3f$ without authentication.

\subparagraph*{Reliable broadcast and support-based communication.}
Byzantine reliable broadcast, introduced by Bracha~\cite{bracha_asynchronous_1987},
ensures consistent message delivery despite a Byzantine sender.
Mechanisms that additionally require messages to be justified by previously received information appear in Coan's compiler~\cite{Coan1988} and implicitly in~\cite{dwork_consensus_1988}.
A related support-based mechanism is used in the simulation of omission failures by Byzantine failures described by Attiya and Welch~\cite{AttiyaW2004}, building on Coan's approach.

Civit et al.~\cite{civit2022crime} give a generic transformation, building on classical crash-to-Byzantine simulations, that uses reliable dissemination together with verification of a message's causal history.  Their transformation relies on authenticated messages; 
in particular, authentication is used to prevent a Byzantine process from fabricating messages in the causal history it reports.
In contrast, our construction does not rely on authentication.

Our supported reliable broadcast delivers a reliably broadcast message to the protocol only when the protocol-specific admissibility predicate holds.
Unlike these earlier applications, which have a simple and largely uniform iterative structure, admissibility in our algorithm depends on the sender's evolving position in the classifier tree and on information accumulated during previous classifier instances.

%% file: preliminaries.tex
\section{Preliminaries}\label{sec:Model and def}

We assume a distributed asynchronous message passing system with $n$ processes, $p_1,p_2,...,p_n$. 
Every pair of processes can send messages to each other
and communication is reliable, that is, 
there is no loss, corruption or creation of messages; 
we assume a process also sends a message to itself. 
A process is \emph{correct} if it follows its protocol;
otherwise, it is \emph{Byzantine} and can deviate from its protocol 
in an arbitrary manner.
We assume that there is a known bound, $f$, on the number of 
processes that can be Byzantine in any execution of the algorithm.
We denote by $C$ the set of correct processes in an execution.

\subparagraph*{Byzantine lattice agreement (BLA).} 
The original definition of \emph{lattice agreement},
stated for non-malicious failures~\cite{attiya_atomic_1998,Attiya_atomicHR1995},
considers $E$, a lattice of values that can be proposed by a process. 
Let $E = (X, \sqsubseteq, \sqcup)$ be a \emph{join semi-lattice}, 
where $X$ is a set of elements with a partial order $\sqsubseteq$, 
and $\sqcup$ a join operation.
Each process $p_i, i \in [n]$ has input $x_i$ from $X$, 
and in the standard definition, 
it has to output a value $y_i \in X$ with the following properties:

\textbf{Comparability}: For all $p_i \in C$ and $p_j 
\in C$, either $y_i \sqsubseteq y_j$ or $y_j \sqsubseteq y_i$.

\textbf{Downward-Validity}: For all $p_i \in C$, $x_i \sqsubseteq y_i$.

\textbf{Upward-Validity}: $\sqcup \{y_i ~ | ~ p_i \in C\} \sqsubseteq \sqcup \{x_i ~ | ~ p_i \in C\}$. 

When failures are Byzantine, upward validity is changed. 
We use the definition in~\cite{di_luna_byzantine_2020}:

\textbf{Byzantine Upward-Validity}: 
$\sqcup \{y_i ~ | ~ p_i \in C\} \sqsubseteq \sqcup(\{x_i ~ | ~ p_i \in C\} \cup B)$, 
where $B \subseteq X$ and $|B| \leq f$. 

\subparagraph*{Time complexity.}
While message delivery is asynchronous, 
without an upper bound on the message delay, 
we adopt the definition of~\cite{AttiyaW2004}
for time complexity in an asynchronous message-passing system.  

We start by defining a \emph{timed execution} as an execution in which 
nondecreasing nonnegative integers (``times'') are assigned to the
events, with no two events by the same process having the same time.
For every timed execution, we consider the interval starting when 
the last correct process calls the BLA procedure and 
ends when the last correct process returns from it.
Then, scale the times so that the maximum time that elapses between 
the sending and receipt of any message between correct processes is 1.  
Finally, the \emph{time complexity} is the maximum, 
over all scaled timed executions, of the length of this interval.

%% file: SRB.tex
\section{Supported (Byzantine) Reliable broadcast (SRB)}
\label{section:SRB}
\subsection{SRB Abstraction and Properties}
The classical notion of \emph{reliable broadcast}~\cite{bracha_asynchronous_1987}
provides a mechanism by which a process can broadcast a message to all other processes.
A hallmark property of reliable broadcast is \emph{non-equivocation}, namely, two correct processes do not deliver different messages, even if the sender is Byzantine.
Importantly, it also provides a \emph{relay} property, namely, 
that if a correct process delivers a message, 
then all correct processes eventually deliver the same message.

We want to further ensure that a message broadcast by a process (even if faulty)
could have been sent according to the protocol.
We call this mechanism \emph{supported reliable broadcast}
(in short, \emph{SRB}).
Operationally, SRB can be viewed as a \emph{support-carrying reliable broadcast},
in which each broadcast message is accompanied by a support justifying it.

Each message is broadcast together with an explicit \emph{support}, 
mainly, a set of messages whose delivery at the sender is a necessary
precondition for sending this message according to the protocol.
This information is used by an \emph{admissibility predicate} 
that checks protocol state that is guaranteed to propagate among correct processes.
The relay property is then leveraged to check the soundness of the support, by waiting to receive this precursor information.

We next state the five properties of supported reliable broadcast. 
The first four properties coincide with those of reliable broadcast; the last property captures the additional support guarantee provided by SRB.
The $tag$ is used to run several broadcast instances,
at different stages of an algorithm, by a given sender $s$.
Note that different senders can send a message with the same tag; that is, the pair sender-tag uniquely identifies an SRB instance.
We remark that two properties, validity and relay, 
\emph{are guaranteed to hold within constant time bounds,} 
and not only eventually, as commonly specified.

The fifth property assumes a \emph{protocol-specific} 
predicate $\textsc{Admis}(m,S,tag,s)$, 
which can be checked locally.
Intuitively, it holds if the support $S$
justifies the message $m$ for broadcast instance $tag$ and $s$
according to the protocol.

\begin{description}
\item[SRB-Validity (timed).]
If the sender $s$ is correct and invokes $\textsf{SRB-broadcast}(m,S,tag,s)$ at time $T$, then every correct process invokes $\textsf{SRB-deliver}(m,S,tag,s)$ by time $T+3$.

\item[SRB-Agreement.]
If two correct processes invoke $\textsf{SRB-deliver}(m_1,S_1,tag,s)$ and $\textsf{SRB-deliver}(m_2,S_2,tag,s)$, for the same $tag$ and $s$, then $m_1 = m_2$ and $S_1 = S_2$.

\item[SRB-Integrity.]
A correct process invokes $\textsf{SRB-deliver}$ at most once
for every $tag$ and $s$. 
Moreover, if a correct process invokes
$\textsf{SRB-deliver}(m,S,tag,s)$,
and the sender $s$ is correct,
then $s$ previously invoked $\textsf{SRB-broadcast}(m,S,tag,s)$.

\item[SRB-Relay (timed).]
If a correct process invokes
$\textsf{SRB-deliver}(m,S,tag,s)$ at time $T$,
then every correct process invokes
$\textsf{SRB-deliver}(m,S,tag,s)$ by time $T+2$.

\item[SRB-Support Admissibility.]
If a correct process $p$ invokes
$\textsf{SRB-deliver}(m,S,tag,s)$, 
then $\textsc{Admis}(m,S,tag,s)$ holds at $p$.
\end{description}

\subsection{SRB Implementation and Proofs}
\label{appendix:SRB}

An implementation appears in Algorithm~\ref{alg:SRB}.
It follows the structure of Bracha's reliable broadcast~\cite{bracha_asynchronous_1987},
using ideas from~\cite[Algorithm~39]{AttiyaW2004} and~\cite{Coan1988}.
The only addition, marked in blue, is to buffer a message $m$ with support $S$ 
and only SRB-deliver it when $m$ is admissible 
(for a specific protocol) given $tag$, sender $s$ and $S$.

\begin{algorithm}[btp]
\caption{Supported Reliable Broadcast (SRB)}
\label{alg:SRB}
\begin{algorithmic}[1]
\small
\Statex \textbf{SRB-broadcast}$(m,S,tag,s)$ 
\Statex $m$: value to broadcast   
        \Comment{in our algorithm, value and group identity}
\Statex $S$: support for $m$  
        \Comment{in our algorithm, support for group identity}
\Statex $tag$: identification of the message 
\Statex $s$: the sender

\Statex         \Comment{Sender $s$}
\State \textbf{send to all} $\textsc{init}(m,S,tag,s)$ 

\Statex
\If{$\textsc{init}(m,S,tag,s)$ has been received and \textsc{echo} not yet sent for $tag$ and $s$}
        \State \textbf{send to all} $\textsc{echo}(m,S,tag,s)$ \label{SRB:echo}
        \EndIf

\Statex
\Statex \textbf{Upon} receiving \textsc{echo}$(m,S,tag,s)$
    \If{\textsc{echo}$(m,S,tag,s)$ received from $\geq n-f$ processes and \textsc{ready} not yet sent for $tag$ and $s$}\label{alg:brb:echo}
        \State \textbf{send to all} \textsc{ready}$(m,S,tag,s)$ \label{line:ready due to echo}
    \EndIf

\Statex

\Statex \textbf{Upon} receiving \textsc{ready}$(m,S,tag,s)$
    \If{\textsc{ready}$(m,S,tag,s)$ received from $\geq n-2f$ processes 
            and \textsc{ready} not yet sent for $tag$ and $s$}
        \State \textbf{send to all} \textsc{ready}$(m,S,tag,s)$
    \EndIf

    \If{\textsc{ready}$(m,S,tag,s)$ received from $\geq n-f$ processes}
    \different{\If{no message with $tag$ and $s$ in buffer}\label{line:first deliver}         \Comment{new part of the algorithm}
    \State  put $(m,S,tag,s)$ in buffer     
    \Comment this event is considered as \textbf{RB-deliver}$(m,S,tag,s)$
    \label{line:rb-deliver}
    \EndIf}
    \EndIf

\Statex

\Statex \different{\textbf{when} $(m,S,tag,s)$ in buffer and $\textsc{Admis}(m,S,tag,s)$ \label{line:admis} 
    \Comment{\textsc{Admis} appears in Algorithm~\ref{alg:clean-admis}} }

    \different{\If{no message has been SRB-delivered for $(s,tag)$} SRB-deliver$(m,S,tag,s)$}  \label{SRB:admis}
    \EndIf
\label{SRB:wait}

\setcounter{SaveAlgLine}{\value{ALG@line}}
\end{algorithmic}
\end{algorithm}

We say that a process \emph{RB-delivers} a message $m$ with support $S$, when it places the message in its buffer (Line~\ref{line:rb-deliver}).
We next prove that Algorithm~\ref{alg:SRB} satisfies the reliable-broadcast properties and timing guarantees
used in the main text (mostly following~\cite[Section~12.7]{AttiyaW2004}). 
Timing properties that rely on admissibility are proved in  Section~\ref{sec:admissibility}.
To reduce clutter, we consider a single invocation of SRB, 
for fixed sender $s$ and $tag$, which may be omitted when they are clear from the context.

\begin{restatable}[RB-Validity, with timing]{lemma}{RBvalid}
\label{lemma:RB validity}
If a correct process SRB-broadcasts at time $T$, 
then all correct processes RB-deliver the message by time $T+3$.
\end{restatable}

\begin{proof} 
Suppose a correct process $p_i$ SRB-broadcasts a message $m$ with support $S$.
Then it sends an appropriate \textsc{init} message to all processes at time $T$. 
Every correct process $p_j$ receives this \textsc{init} message by time $T+1$.
Since this is the first (and only) \textsc{init} message it receives from $p_i$, 
$p_j$ sends the corresponding \textsc{echo} message to all processes by time $T+1$.

Every correct process $p_j$ receives at least $n-f$
\textsc{echo} messages for $m$ and $S$, 
by time $T+2$.
So $p_j$ receives at most $f$ \textsc{echo} messages
for any $m' \ne m$ or $S' \neq S$.
Since $f < n-f$ (recall $n > 3f$),
$p_j$ sends \textsc{ready} for $m$ and $S$, 
by time $T+2$.

Every correct process $p_j$ receives at least $n-f$
\textsc{ready} messages for $m$ and $S$, 
by time $T+3$.
So $p_j$ receives at most $f < n-2f$ (recall $n > 3f$) \textsc{ready} messages for any $m' \ne m$ or $S' \neq S$,
and RB-delivers $m$ and $S$, by time $T+3$.
\end{proof}

The next lemma is the key to proving SRB-agreement and SRB-relay. 

\begin{lemma}
\label{lemma:same ready}
If one correct process sends a \textsc{ready} message with $m$ and $S$ and another correct process sends a \textsc{ready} message with $m'$ and $S'$, then $m = m'$ and $S = S'$.
\end{lemma}

\begin{proof}
Assume, by way of contradiction, that $m \ne m'$ or $S \ne S'$.
Let $p_i$ be the first correct process to send a \textsc{ready} message for $m$ and $S$.
Since $p_i$ can only receive \textsc{ready} messages from at most $f$ 
faulty processes up till this point and $f$ is less than $n-2f$,
$p_i$ sends \textsc{ready} in Line~\ref{line:ready due to echo}.
That is, $p_i$ received at least $n-f$ \textsc{echo} messages
for $m$ and $S$,
and at least $n-f-t$ of these are from correct processes,
where $t$ is the actual number of faulty processes in the execution.

Similarly, $p_j$, the first correct process to send
\textsc{ready} message for $m'$ and $S'$,
received at least $n-f-t$ \textsc{echo} messages
for $m'$ and $S'$ from correct processes.

Since each correct process sends only one echo message 
(for a given sender and tag), 
the total number of correct processes is at least $2(n-f-t)$.
We have $2(n-f-t)+t \leq n$, 
so $n\leq 2f+t\leq 3f$ (since $t \leq f$), 
which is a contradiction.
\end{proof}

\begin{restatable}{lemma}{SRBagree}
\label{lemma:SRB agreement}
Algorithm~\ref{alg:SRB} satisfies SRB-agreement.
\end{restatable}

\begin{proof}
Suppose correct process $p_i$ delivers $m$ and $S$,
while a correct process $p_j$ delivers $m'$ and $S'$.
Assume for contradiction that either $m \ne m'$ or $S \ne S'$.
Thus $p_i$ receives at least $n-f$ \textsc{ready} messages
for $m$ and $S$, at least $n-2f > 0$ of which are from correct processes.
Similarly, $p_j$ receives at least $n-f$ \textsc{ready} messages
for $m'$ and $S'$, at least $n-2f > 0$ of which are from correct processes.
But this violates Lemma~\ref{lemma:same ready}.
\end{proof}

\begin{restatable}{lemma}{SRBintegrity}
\label{lemma:SRB integrity}
Algorithm~\ref{alg:SRB} satisfies SRB-integrity.
\end{restatable}

\begin{proof}
The first part (no duplicates) is ensured by Line~\ref{line:first deliver}, 
which avoids buffering two different messages for the same tag,
hence, two different messages cannot be delivered for the same tag, 
in Line~\ref{SRB:admis}.

Suppose correct process $p_i$ delivers $m$ and $S$.
Thus $p_i$ receives at least $n-f$ \textsc{ready} messages
for $m$ and $S$, at least $n-2f$ of which are from correct processes.
Let $p_h$ be the first correct process to send \textsc{ready} for $m$ and $S$.
Then $p_h$ cannot have received $n-2f$ \textsc{ready} messages already,
since $n-2f > f$ (recall that $n>3f$) and up till now only
faulty processes have sent \textsc{ready} for $m$ and $S$.
Thus $p_h$ receives at least $n-f$ \textsc{echo} messages for $m$ and $S$,
at least $n-2f$ of which are from correct processes.
A correct process only sends an \textsc{echo} message for $m$ and $S$,
if it receives \textsc{init} messages for $m$ and $S$ from the sender,
as needed. 
\end{proof}

\begin{restatable}[RB-Relay, with timing]{lemma}{RBrelay}
\label{lemma:RB relay}
If some correct process RB-delivers $m$ and $S$ at time $T$, 
then every correct process RB-delivers $m$ and $S$ by time $T+2$.
\end{restatable}

\begin{proof} 
Suppose correct process $p_i$ RB-delivers $m$ and $S$ at time $T$.
Then $p_i$ receives at least $n-f$ \textsc{ready} messages for $m$ and $S$,
at least $n-2f$ of which are from correct processes.
Thus every correct process $p_j$ receives at least $n-2f$
\textsc{ready} messages for $m$ and $S$, by time $T+1$.
By Lemma~\ref{lemma:same ready}, 
$p_j$ does not send a \textsc{ready} message for $m' \neq m$ or $S' \neq S$, and therefore, 
$p_j$ sends \textsc{ready} messages for $m$ and $S$, by time $T+1$.
Thus every correct process receives at least $n-f$
\textsc{ready} messages for $m$ and $S$ by time $T+2$,  
and RB-delivers $m$ and $S$ (Line~\ref{line:rb-deliver}).
\end{proof}

By the code (Line~\ref{SRB:admis}):

\begin{proposition}\label{prop:SRB support admissibility}
Algorithm~\ref{alg:SRB} satisfies SRB-Support Admissibility.
\end{proposition}

%% file: one-shot-BLA.tex
\section{One-Shot Byzantine Lattice Agreement}
\label{section:BLA}

\subsection{Overview}

The overall structure of the algorithm follows the lattice agreement
of Attiya and Rachman~\cite{attiya_atomic_1998};
this is a wait-free algorithm for asynchronous shared-memory,
and it has $O(n \log n)$ step complexity.
The algorithm internally operates on sets of \emph{process-value pairs}, ordered by inclusion; at termination, each process returns the join of the values in its final set.
These sets are ordered by inclusion, with union as their join.

For the description below and the proof that follows, 
let $f'$ be the smallest power of 2 that is greater than or equal to $f$, 
and denote $F = \log f' +2$; 
note that $f\le f'<2f$ and that $F$ is an integer in $O(\log f)$.

The algorithm is based on a sub-construct, called a \emph{classifier},
parameterized by a threshold $K$.
Given a collection of sets (one per process),
a classifier partitions the processes that access it into two groups.
Processes in the first group, called \emph{low-information}
(in short \emph{low}), retain their input sets, 
whose size is at most $K$.
Processes in the second group, called \emph{high-information}
(in short \emph{high}), obtain a set of size greater than $K$ that contains the union of all sets held by processes in the first group.

\begin{figure}
\centering
\includegraphics[scale=.45]{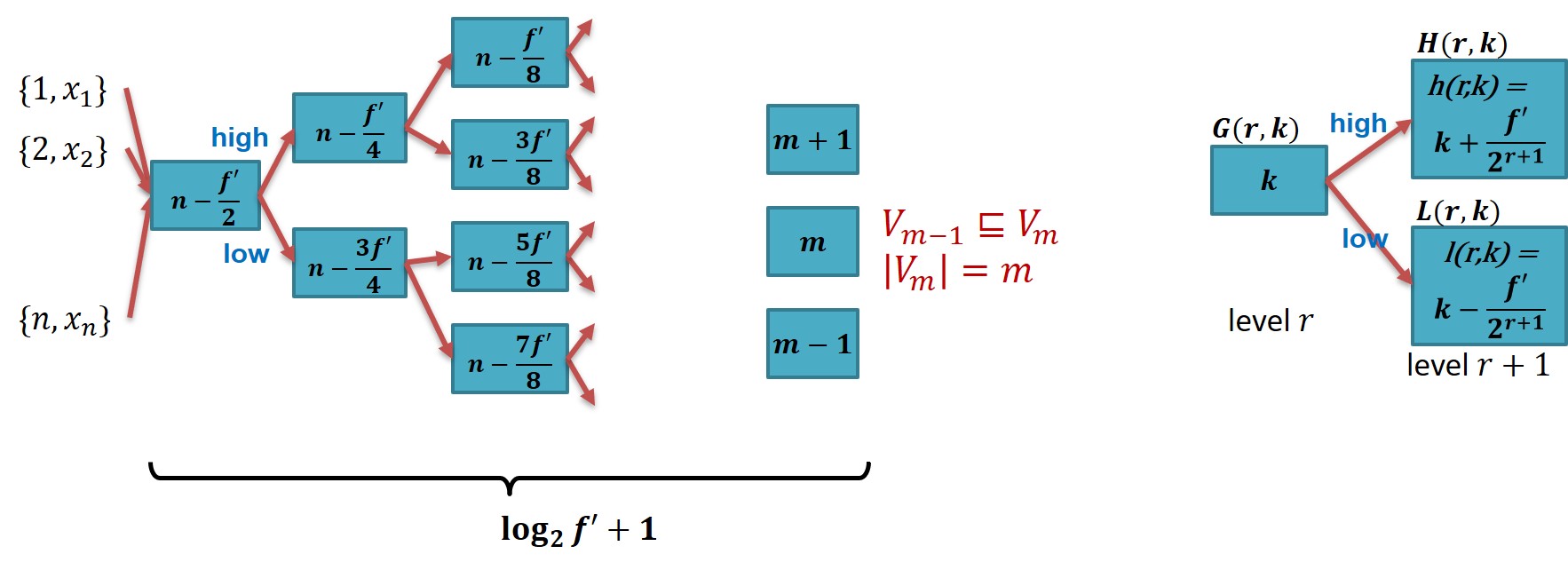}  
\caption{Left-hand side: the tree of classifiers used in our lattice agreement algorithm.
Right-hand side: zoom-in on a classifier at level $r$ and its two children.}
\label{figure:classifiers tree}
\end{figure}

Classifiers are organized in a full binary tree of logarithmic depth,
with carefully chosen parameters (see Figure~\ref{figure:classifiers tree}).
Each process follows a root-to-leaf path in this tree.
At each node, a process accesses the classifier and moves to the down
or up child according to whether it is classified as low or high.
Along any path, sets only grow, and whenever processes split at a node
(illustrated in Figure~\ref{figure:classifiers tree}),
the set of a process that moves up contains the set of a process that
moves down.
Consequently, it can be shown that the sets returned by all processes
at the leaves are totally ordered by inclusion.

There are two main challenges when translating this algorithm to the message-passing setting with Byzantine failures.

The first is to replace reads and writes with message-passing communication;
this challenge is resolved by interacting through quorums,
in a known manner that originated in~\cite{AttiyaBD1995}.

The second, more difficult, challenge is to deal with the possibility of
Byzantine failures.
Part of this challenge can be dealt with using known tools, 
such as reliable broadcast~\cite{bracha_asynchronous_1987}.
However, the way information flows in the classifier tree complicates matters:
a Byzantine process that moves down (as a low process) may introduce a totally
new value that does not appear in the sets of the high processes that move up.
In the shared-memory setting, this cannot happen because writes are immediately
visible; in a message-passing system with Byzantine failures, 
this implicit assumption no longer holds.

We address this challenge by using \emph{supported reliable broadcast},
which requires processes to show, not just tell:
any value introduced by a process is accompanied by a support showing that it
could have been sent according to the protocol.

\subsection{Message Formats and Local Variables} \label{appendix:messages-variables}

\subparagraph*{Message types for the classifier.}

Below, $\mathcal{U}$ is the set of all process-value pairs.

\begin{itemize}
  \item \textbf{Init message} has tag \textsf{init}, 
  carries the sender's input pair $(i,x_i)$, and has empty support.
  It disseminates the sender's initial value starting the classifier tree.

  \item \textbf{Write message}
  has tag $(\textsf{write},r)$ and carries 
  a set of values $V$, \emph{class} and label $k$ and support $Sup$.
  It represents a process's write at instance $(r,k)$ together with information justifying that this write could have been issued 
  according to the protocol, where $class\in\{\textsf{low},\textsf{high}\}$ 
  indicates the sender's classification at this instance.

  \item \textbf{Write acknowledgment}
  has tag $(\textsf{ackW},r,j)$ and carries a set of values $V$.
  It acknowledges the delivery of process $p_j$'s \textsf{write} at level $r$, where $V$ is the sender's locally known values.

  \item \textbf{Read message}
  has tag $(\textsf{read},r)$ and carries the label $k$ and an array mapping each process to a write-acknowledgment; 
  these correspond to values received by the sender in the write phase of the classifier. 

  \item \textbf{Read acknowledgment}
  has tag $(\textsf{ackR},r,j)$ and carries the set $V$ of values known to the sender and the label $k$.
  It acknowledges the delivery of process $p_j$'s \textsf{read} at level $r$.

  \item \textbf{High-read request}
  has tag $(\textsf{hreq},r)$ and carries $V$ the 
  set of values previously obtained and the label $k$;
  it is used by a process classified as \emph{high} at this instance to perform an additional collect step.

  \item \textbf{High-read response} 
  has tag $(\textsf{hrep},r,j)$ and carries a set $V$,
  which are the sender's locally known values, 
  and the label $k$,
  in response to a \textsf{hreq} issued by process $p_j$.
\end{itemize}

\subparagraph*{Local variables for a classifier instance.}

\begin{itemize}
  \item \textbf{Known values.}
  $Knw_i[r,k] \subseteq \mathcal{U}$, initially $\emptyset$.
  The set of values that $p_i$ has learned at instance $(r,k)$,
  namely the union of all value sets carried by \textsf{write} messages
  that $p_i$ has SRB-delivered.

  \item \textbf{Write acknowledgments.}
  $ackW_i[r,j,h] \in 2^{\mathcal{U}} \cup \{\emptyset\}$, initially $\emptyset$.
  This variable stores the value sent in an \textsf{ackW} message sent by process $p_h$ acknowledging the \textsf{write} issued by process $p_j$ at level $r$.

  \item \textbf{Read acknowledgments.}
  $ackR_i[r,j,h] \in 2^{\mathcal{U}} \cup \{\emptyset\}$, initially $\emptyset$.
  This variable stores the value sent in an \textsf{ackR} message sent by process $p_h$ acknowledging the \textsf{read} issued by process $p_j$ at level $r$.

  \item \textbf{Support for the next level.}
  $Sup_i[r]$, initially $\emptyset$.
  This variable stores the support that $p_i$ will attach to its next-level
  \textsf{write}.
  It is derived from the locally stored read acknowledgments  $ackR_i[r,i,\cdot]$ and consists of items of the form $\langle h,R\rangle$, where $R$ was received from $p_h$.

  \item \textbf{Recorded writes.}
  $Lwr_i[r,k,j] \in 2^{\mathcal{U}} \cup \{\emptyset\}$, initially $\emptyset$.
  This variable records the value set $V$ written by process $p_j$ when at instance $(r,k)$.
  This information is later consulted by the admissibility predicate to ensure that a process classified as low does not introduce new values at subsequent levels.

  \item \textbf{Recorded classifier.} $Lab_i[r,j]\in[0,n]\cup\{\bot\}$, initially $\bot$.
  This variable records the label carried by the unique write message
of process $p_j$ at level $r$ that is SRB-delivered at $p_i$.
  Subsequent admissibility checks use this recorded label to ensure
that messages sent by $p_j$ at level $r$ are consistent with the
classifier identified by its write.

  \item \textbf{High-read responses.}
  $Hrp_i[r,j,h] \in 2^{\mathcal{U}} \cup \{\emptyset\}$, initially $\emptyset$.
  This variable stores the value of the \textsf{hrep} message
  sent by process $p_h$ in response to a high-read request issued by process $p_j$ at level $r$. 
\end{itemize}

\subsection{The Algorithm}

Algorithm~\ref{alg:main} shows the main code, which closely follows 
the structure of Attiya-Rachman, funneling processes through a sequence of classifiers with carefully chosen labels. 

Unlike the original shared-memory algorithm,
where each process starts the classifier tree with a singleton set
containing only its own value, 
our message-passing Byzantine setting uses an initial dissemination phase.
Starting the classifier tree with a singleton set would still yield 
a correct algorithm, but would result in $O(\log n)$ levels.
Instead, each process first uses SRB to collect a set of 
at least $n-f$ process-value pairs.
This ensures correct processes enter the
classifier tree with a sufficiently large set, 
allowing the depth of the tree to be in $O(\log f)$.
The use of SRB also prohibits Byzantine processes from introducing 
new values to some of the processes, later in the algorithm.
Only after this initial phase do processes enter the classifier tree
and repeatedly invoke the classifier procedure.

\begin{algorithm*}[tbp]
\small
\caption{Logarithmic algorithm for Byzantine lattice agreement: code for process $p_i$}
\label{alg:main}
\begin{algorithmic}[1]
   \setcounter{ALG@line}{\value{SaveAlgLine}}

\Statex \textbf{Local variables}
\State $x_i$: input value; $y_i$: output value
\State $l_i$: label, initially $n - \frac{f'}{2}$ 
\State $class_i$: classification, initially $low$
\State $V_{i,r}$: set of values at level $r = 1, \ldots, F$ 

\Statex 
\State \textsf{SRB-broadcast}$((i,x_i),\emptyset,\textsf{init},i)$ \label{MA:send-Init} 
\Comment{level 0}
\State \textbf{wait until} $Knw_{i}[0,n]$ contains at least $n-f$ pairs, including $(i,x_i)$
        \label{MA:Init-1} 
\State $V_{i,1} \gets Knw_{i}[0,n]$

\Statex  
\For{$r = 1$ to $F-1$} \label{MA:Loop-Start} 
        \Comment{levels $1$ to $F-1$}
    \State $(V_{i,r+1}, class_i) \gets \textsc{Classifier}(class_i,V_{i,r},  r,l_i)$ \label{MA:Loop-1}

    \If{$class_i = high$} \label{MA:Loop-2}
        $l_i \gets l_i + f'/2^{r+1}$ \label{MA:Loop-3}
    \Else { $l_i \gets l_i - f'/2^{r+1}$} \label{MA:Loop-4}
    \EndIf

\EndFor \label{MA:Loop-End}

\State $y_i \gets \sqcup \{ x \mid (j,x) \in V_{i,F} \}$ 
\label{MA:Output}

\Statex
\Statex \textbf{Upon} SRB-deliver$((j,x_j),\emptyset,\textsf{init},j)$ 
\label{MA:DeliverX-1}
\State \quad $Knw_{i}[0,n] \gets Knw_{i}[0,n] \cup \{(j,x_j)\}$
\label{MA:DeliverX-2}
\State \quad $Lab_{i}[0,j]  \gets n$ 
\setcounter{SaveAlgLine}{\value{ALG@line}}
\end{algorithmic}
\end{algorithm*}

The \textsf{Classifier} procedure appears in Algorithm~\ref{alg:clean-classifier},
together with code describing how messages are handled.
Each classifier instance is identified by a level $r$ and a label $k$.
Note that the label of a classifier also indicates its level;
that is, a label at level $r$ has the form
$n- a\frac{f'}{2^r}$, where $a$ is odd.
In particular, labels do not repeat between the levels. 
For clarity, however, we still index local arrays and tag messages 
both by their level ($r$) and label ($k$). 

\begin{algorithm}[tb]
\small
\caption{Classifier for process $p_i$}
\label{alg:clean-classifier}
\begin{algorithmic}[1]
   \setcounter{ALG@line}{\value{SaveAlgLine}}
   
\Function{Classifier}{$class,V,r,k$}

\State $\textsf{SRB-broadcast}((V,k,class),Sup_i[r-1],(\textsf{write},r),i)$
    \Comment{write value} 
    \label{NCL:WriteStep-4}
 
\State \textbf{wait until }
$|\{h : ackW_i[r,i,h]\neq\emptyset\}|\ge n-f \wedge Lwr_i[r,k,i]\neq\emptyset$  
    \label{NCL:WriteStep-5}

\State $\textsf{SRB-broadcast}(k,ackW_i[r,i,\cdot],(\textsf{read},r),i)$
    \Comment{first collect}
    \label{NCL:1collect-1}
\State \textbf{wait until} $|\{h : ackR_i[r,i,h]\neq\emptyset\}|\ge n-f$
    \label{NCL:1collect-2}
\State $Sup_i[r] \gets \{\langle h,ackR_i[r,i,h]\rangle : ackR_i[r,i,h]\neq\emptyset\}$ \label{NCL:1collect-3}

\State $\textit{ReadValues} \gets \cup_h    ackR_i[r,i,h]$
    \label{NCL:1collect-4}

  \If{$|\textit{ReadValues}| > k$}
     \State $\textsf{SRB-broadcast}(k,\textit{ReadValues},(hreq,r),i)$
            \Comment{second collect for high processes}
            \label{NCL:Classify-2}
     \State \textbf{wait until} $\left|\{h : Hrp_i[r,i,h] \neq \emptyset\}\right| \ge n-f$  \label{NCL:Classify-3}
     \State \Return $\left(\cup_h  Hrp_i[r,i,h], high\right)$ \label{NCL:Classify-4}
  \Else \ \Return $(V, low)$ \label{NCL:Classify-5}
  \EndIf
\EndFunction

\Statex \Comment{Handlers for the classifier}
\Statex \textbf{Upon} $\textsf{SRB-deliver}((H,k,class),Sup,(write,r),h)$:
  \State\quad $Knw_i[r,k] \gets Knw_i[r,k] \cup H$
                \label{NCL:DeliverWrite-2}
   \State\quad $Lwr_i[r,k,h] \gets H$ \label{NCL:DeliverWrite-L}
   \State\quad  $Lab_i[r,h] \gets k$ \label{NCL:DeliverWrite-Lab}
  \State\quad $\textsf{SRB-broadcast}((Knw_i[r,k],k),\emptyset,(ackW,r,h),i)$ \label{NCL:DeliverWrite-4}

\Statex \textbf{Upon} $\textsf{SRB-deliver}((H,k),\emptyset,(ackW,r,j),h)$:
  \State\quad $ackW_i[r,j,h] \gets H$ \label{NCL:DeliverAckWrite}

\Statex \textbf{Upon} $\textsf{SRB-deliver}(k,Sup,(read,r),h)$:
  \State\quad $\textsf{SRB-broadcast}((Knw_i[r,k],k),\emptyset,(ackR,r,h),i)$ \label{NCL:DeliverRead-4}

\Statex \textbf{Upon} $\textsf{SRB-deliver}((H,k),\emptyset,(ackR,r,j),h)$:
  \State\quad $ackR_i[r,j,h] \gets H$

\Statex \textbf{Upon} $\textsf{SRB-deliver}(k,Sup,(hreq,r),h)$:
  \State\quad $\textsf{SRB-broadcast}((Knw_i[r,k],k),\emptyset,(hrep,r,h),i)$
  \label{NCL:DeliverHigh-3}

\Statex \textbf{Upon} $\textsf{SRB-deliver}((H,k),\emptyset,(hrep,r,j),h)$:
  \State\quad $Hrp_i[r,j,h] \gets H$ 
\setcounter{SaveAlgLine}{\value{ALG@line}}

\end{algorithmic}
\end{algorithm}

The classifier is executed with a set $V$ of values and a classification 
(either \emph{low} or \emph{high}) inherited from the previous level.
The classifier partitions processes into low- and high-information groups 
while enforcing controlled information flow, even in the presence of Byzantine failures.

Classifier communication exclusively uses supported reliable broadcast (SRB).
Each message carries a tag identifying its type and the classifier instance, in addition to a value and a support. 
Recall that messages are SRB-delivered only if they satisfy the admissibility predicate \textsc{Admis}, explained later.
Thus, a message is SRB-delivered only after its protocol-specific admissibility condition holds at the receiver.

\subparagraph*{The classifier's control flow.}
First, a process $p_i$ SRB-broadcasts its current set $V$ using a \textsf{write} message, attaching the support $Sup_i[r-1]$,
and waits for \textsf{ackW} responses, which are stored in $\textsf{ackW}_i$.
Upon delivering \textsf{write} messages, 
processes update their accumulated sets and send \textsf{ackW} acknowledgments.
Next, $p_i$ performs a collect by SRB-broadcasting a \textsf{read}
message containing its write acknowledgments and waits for $n-f$
\textsf{ackR} responses, which are stored in $\textsf{ackR}_i$. 
Finally, the process computes the union of the collected values and classifies itself as high if the union exceeds $k$; in this case, it performs an additional collect by SRB-broadcasting a \textsf{hreq} message and waiting for $n-f$ \textsf{hrep} messages.
Otherwise, the process is classified as low and retains its set $V$.
The second collect is needed to ensure that the sets of high processes contain the sets of low processes
(similarly to~\cite{attiya_atomic_1998}).

A \textsf{write} message carries the sender's current set, label, and classification, 
together with support for the transition from the previous classifier.
The \textsf{read}/\textsf{ackR} messages implement the first collect, 
while \textsf{hreq}/\textsf{hrep} implement the additional collect performed by high processes.

The label of a classifier is deliberately not carried in the SRB tag.
Since the tag of a write message at level~$r$ is just $(\textsf{write},r)$, 
all writes of a given process at level $r$ must belong to the same SRB instance.  
SRB-Integrity and SRB-Agreement therefore prevent a Byzantine process from having different writes delivered for different classifiers at the same level.
The unique label carried by this write is recorded and 
is checked against the labels carried by subsequent messages.

Similarly, a process has a single SRB instance for its
\textsf{read} and \textsf{hreq} messages at each level.
For reply messages, the process being answered is included in the tag, 
so there is one SRB instance for each sender, level, reply type, and requester.

For each classifier instance identified by a level $r$ and label $k$, 
process $p_i$ maintains $Knw_i[r,k]$, the values learned from delivered writes;
$Lwr_i[r,k,j]$ and $Lab_i[r,j]$, recording the delivered write and classifier label of $p_j$; 
and the acknowledgments and high-read responses received during the classifier.
The set $Sup_i[r]$ consists of the read acknowledgments used to justify $p_i$'s next low write.

All set-valued variables are monotone.
The remaining variables are initially empty or undefined and, once set, are never changed, because each process has a unique SRB-delivered message of the corresponding type.

\subparagraph*{The admissibility predicate.}
Algorithm~\ref{alg:clean-admis} describes the protocol-specific admissibility 
procedure we use to ensure that every SRB-delivered message is
consistent with a possible execution of the classifier by its sender.
\textsc{Admis} checks both the information carried by a message
and the sender's claimed position in the classifier tree.

A \textsf{write} message at level $r$ must be consistent with the sender's state at its
parent classifier at level $r-1$.
If the sender is high, its new set may have grown, but every value in it
must already be known at the parent classifier, and the label of the parent
must be the one leading to the claimed high child.
If the sender is low, its set must remain unchanged from its write at the
parent classifier.
Moreover, its support consists of at least $n-f$ \textsf{ackR} messages
from the preceding read phase; their union must contain the written set and
have size at most the parent threshold.
Thus, a process claiming to move to the low child cannot subsequently add
information that was not exposed during the parent classifier.
At the first level there is no parent classifier, so a low write is instead
required to contain at least $n-f$ values already obtained in the initial
dissemination phase.

A \textsf{read} is admissible if it is supported by at least $n-f$ \textsf{ackW} messages
indicating that the corresponding process has performed the \textsf{write} of the same level.

The remaining checks ensure that communication within a classifier is
consistent with the recorded classifier state.
A high-read request may contain only values already known at the classifier.
Finally, an \textsf{ackW}, \textsf{ackR}, or \textsf{hrep} response may
contain only values known at the indicated classifier, and is accepted only
after the classifier label of the process being answered has been established.
The \textsf{init} case simply checks that the sender introduces a single
value under its own identity.

\textsc{Admis} is deliberately local and evaluates the message and support relative to information previously SRB-delivered at the receiving process.
The crucial argument is not only that a correct process satisfies the
predicate when it sends a message---this follows directly from the protocol
and is formalized in Lemma~\ref{lemma:local-admissibility}---but that information
needed to establish admissibility for all correct processes propagates sufficiently quickly to them, as shown in Lemma~\ref{lemma:propagation}.

\begin{algorithm}[tb]
\small
\caption{Admissibility checks for the classifier}
\label{alg:clean-admis}
\begin{algorithmic}[1]
\setcounter{ALG@line}{\value{SaveAlgLine}}

\Function{ConfRead}{$A,j,r,k$}

\If{$|\{h : A[h]\neq\emptyset\}| < n-f$
      $\vee$ $Lwr_i[r,k,j]=\emptyset$}
    \Return false
    \label{confRead-check-write}
  \EndIf
  \ForAll{$h$ with $A[h]\neq\emptyset$}
    \If{$A[h]\neq ackW_i[r,j,h]$} \Return false \EndIf
    \label{confRead-check-ackW}
  \EndFor
  \State \Return true
\EndFunction

\Statex
\Function{ConfSup}{$Sup,j,r$}
  \Comment{$Sup$ is a set of proof items $\langle h,R\rangle$}
  \If{$|Sup| < n-f$} \Return false \EndIf
  \ForAll{$\langle h,R\rangle \in Sup$}
     \If{$ackR_i[r,j,h]\neq R \vee R= \emptyset$} \Return false \EndIf
  \EndFor
  \State \Return true
\EndFunction

\Statex
\Function{Admis}{$P,Sup,tag,j$}
  \If{$tag = (\textsf{write},r)$} 
  \Comment{$P$ is a set-label-class tuple}
  \State $(H,k,class) \gets P$ ; $r' \gets r-1$
  \If{$class = high$} 
     \State $k' \gets k-f'/2^{r}$  
     \State \Return ($(r>1) \wedge\ Lab_i[r',j]=k'
     \wedge\ (H\subseteq Knw_i[r',k'])$) 
     \label{NCL:ValidFunc-3}
  \EndIf
  \If{$class = low$}
  \State $k' \gets k+f'/2^{r}$  
  \If{$r=1$} \Return ($ H \subseteq Knw_i[r',k'] \wedge |H|\geq n - f$) \label{round1write}
   \EndIf
   \State \Return $\Big(
        H  \subseteq Knw_i[r',k'] 
        \ \wedge\ 
        \textsc{ConfSup}(Sup, j, r')
        \ \wedge\
        H \subseteq \bigcup\{R : (-,R) \in Sup\}
        \ \wedge\ $
        \Statex \quad \quad \quad \quad \quad \quad \quad 
        $\left|\bigcup\{R : (-,R) \in Sup\}\right| \le k'
        \wedge
        H = Lwr_i[r',k',j] \Big)$ \label{NCL:ValidFunc-2}
  \EndIf
  \EndIf
  
  \If{$tag = (\textsf{ackR},r,h) \vee tag = (\textsf{ackW},r,h) \vee tag = (\textsf{hrep},r,h)$} 
  \Comment{$P$ is a set-label pair}
  \State $(H,k) \gets P$
  \State \Return $H \subseteq Knw_i[r,k] ~\wedge~ Lab_i[r,h] = k$  \label{NCL:ValidFunc-5}
  \EndIf

  \If{$tag = (\textsf{hreq},r)$}  \Comment{in all remaining cases, $P$ is only the label}
  \State
     \Return $Sup \subseteq Knw_i[r,P] ~\wedge~ Lab_i[r,j] = P$   \label{ackW}
  \EndIf

  \If{$tag = (\textsf{read},r)$}
     \Return \textsc{ConfRead}$(Sup,j,r,P) ~\wedge~ Lab_i[r,j] = P$ \label{NCL:ValidFunc-4}
  \EndIf

  \If{$tag = \textsf{init}$}
    \Return ($P=(j,x_j) \wedge \ x_j \in X$) \label{NCL:ValidFunc-6}
  \EndIf
  \State \Return false \Comment{ignore ill-formed messages}
\EndFunction
\setcounter{SaveAlgLine}{\value{ALG@line}}

\end{algorithmic}
\end{algorithm}

\section{Correctness proof}

We start by proving the correctness and timing properties of the admissibility checks,
then, we prove the classifier properties,
and finally, we put all the pieces together
to prove the correctness and complexity of the algorithm.

We use a simple invariant, implied by the fact that 
$\mathit{Knw}_i[r,k]$ is only updated in Line~\ref{NCL:DeliverWrite-2}:

\begin{invariant}\label{invariant:know monotonicity}
For every correct process $p_i$, every level $r \ge 0$, 
and every label $k$, 
whenever a message $((V,k,class),\mathit{Sup},(\mathsf{write},r))$ is
SRB-delivered by $p_i$, $V$ is added to $\mathit{Knw}_i[r,k]$.
Consequently, $\mathit{Knw}_i[r,k]$ is monotone.
\end{invariant}

\subsection{Admissibility} 
\label{sec:admissibility}

The admissibility predicate is evaluated with respect to 
the local state of the process invoking it. 
By Invariant~\ref{invariant:know monotonicity},
every condition involving $Knw$ is preserved.
The other local variables inspected by \textsc{Admis} are write-once,
while all remaining conditions depend only on the message and its support. 
This implies a \emph{monotonicity property}:
if $\textsc{Admis}(P,S,\mathit{tag},j)$ 
holds at a correct process $p_i$ at time $T$, 
then it holds at $p_i$ at every time $> T$.

We first prove that a correct process $p_i$ SRB-broadcasts a message $P$ with support $S$, 
only when $P$ is admissible (as evidenced by $S$) at $p_i$. 
When it is clear from the context, 
we sometimes omit the sender from the arguments of the message
or the function \textsc{Admis}.

\begin{lemma}\label{lemma:local-admissibility}
  If a correct process $p_i$ SRB-broadcasts $(P,S,\mathit{tag})$, 
then $\textsc{Admis}(P,S,\mathit{tag},i)$ holds at $p_i$.
\end{lemma}

\begin{proof}
We proceed by case analysis, according to the message type.
\begin{enumerate}
\item 
$p_i$ broadcasts $((i,x_i),\emptyset,\textsf{init})$ (Line~\ref{MA:send-Init}): 
Since $x_i\in X$, 
the condition in Line~\ref{NCL:ValidFunc-6} holds.

\item $p_i$ broadcasts
$((H,k,\textsf{high}),Sup_i[r-1],(\textsf{write},r))$:
Denote $r'=r-1$ and $k'=k-\frac{f'}{2^r}$.
Since the classifier for level 1 is called with \textsf{low},
it follows that $r>1$.
Moreover, $H$ is the value returned by the high branch of the
classifier at level $r'$, namely,
$H=\bigcup_h Hrp_i[r',i,h]$. 
Every nonempty entry $Hrp_i[r',i,h]$ was obtained by
SRB-delivering a message $((H_h,k'),\emptyset,(\textsf{hrep},r',i))$ from process $p_h$.
By SRB-Support Admissibility
(Proposition~\ref{prop:SRB support admissibility}) and
Line~\ref{NCL:ValidFunc-5}, 
$H_h\subseteq Knw_i[r',k']$ and $Lab_i[r',i]=k'$.
Therefore, 
$H \subseteq Knw_i[r',k']$ and $Lab_i[r',i]=k'$; 
and the condition in Line~\ref{NCL:ValidFunc-3} holds.

\item $p_i$ broadcasts 
$((H,k,\textsf{low}),Sup_i[r-1],(\textsf{write},r))$. 
If $r=1$, then $k=n-\frac{f'}{2}$ and $H = V_{i,1} \subseteq Knw_i[0,n]$.
By Line~\ref{MA:Init-1}, $|H|\ge n-f$, 
and the condition in Line~\ref{round1write} holds.

Assume now that $r>1$, and let
$r'=r-1$ and $k'=k+\frac{f'}{2^r}$. 
Since $p_i$ was classified low at level $r'$, 
it retains its input to that classifier. 
Thus, $H$ is also the set written by $p_i$ at level $r'$. 
By the second condition in Line~\ref{NCL:WriteStep-5}, 
this write has been SRB-delivered at $p_i$, and therefore,
$H=Lwr_i[r',k',i]$ and $H\subseteq Knw_i[r',k']$. 
Furthermore,
$Sup_i[r'] =
  \{\langle h,ackR_i[r',i,h]\rangle:
       ackR_i[r',i,h]\neq\emptyset\}$,
and Line~\ref{NCL:1collect-2} ensures that
$|Sup_i[r']|\ge n-f$,
and $\Call{ConfSup}{Sup_i[r'],i,r'}$ holds.

Moreover, for each $ackR$ message in $Sup_i[r']$ sent by a correct process $p_h$ 
(there is at least one since $|Sup_i[r']|\ge n-f > f$), 
$p_h$ previously verified the admissibility of the message with tag $(read,r')$ by $p_i$. 
In particular,$p_i$ verified Line~\ref{confRead-check-write},
so $H\subseteq Knw_h[r',k']$ before $p_h$ 
$\textsf{SRB-broadcast}((Knw_h[r',k'],k'),\emptyset,(ackR,r',i))$. 
Hence,
$H\subseteq \bigcup\{R:\langle -,R\rangle\in Sup_i[r']\}$.
$H\subseteq \bigcup\{R:\langle -,R\rangle\in Sup_i[r']\}$
$\subseteq$\textit{$ReadValues$} 
computed by $p_i$ at level $r'$. 
Since $p_i$ is classified as low at level $r'$, 
$\left| \textit{ReadValues} \right|\le k'$,
and the condition in Line~\ref{NCL:ValidFunc-2} holds.

\item 
$p_i$ broadcasts
$(k,ackW_i[r,i,\cdot],(\textsf{read},r))$,
after waiting that 
$|\{h:ackW_i[r,i,h]\neq\emptyset\}|\ge n-f$ 
and $Lwr_i[r,k,i]\neq\emptyset$. 
Since the support is $ackW_i[r,i,\cdot]$, 
all the conditions of $\Call{ConfRead}{ackW_i[r,i,\cdot],i,r,k}$ hold. 
Moreover, the write delivery that sets
$Lwr_i[r,k,i]$ also sets $Lab_i[r,i]=k$,
and Line~\ref{NCL:ValidFunc-4} holds.

\item 
$p_i$ broadcasts
$(k,\textit{ReadValues},(\textsf{hreq},r))$ 
and $\textit{ReadValues}=\bigcup_h ackR_i[r,i,h]$.
A nonempty entry $ackR_i[r,i,h]$ was obtained by
SRB-delivering a message 
$((H_h,k),\emptyset,(\textsf{ackR},r,i))$.
By SRB-Support Admissibility
(Proposition~\ref{prop:SRB support admissibility}) and Line~\ref{NCL:ValidFunc-5},
$H_h\subseteq Knw_i[r,k]$ and $Lab_i[r,i]=k$. 
By Invariant~\ref{invariant:know monotonicity}, 
$\textit{ReadValues} \subseteq Knw_i[r,k]$;
since $Lab_i[r,i]=k$, the condition of Line~\ref{ackW} holds.

\item 
$p_i$ broadcasts a message $((Knw_i[r,k],k),\emptyset,tag)$ 
with $tag$ 
$(\textsf{ackW},r,h)$, or $(\textsf{ackR},r,h)$,
or $(\textsf{hrep},r,h)$.
In all three cases, 
the message is broadcast only after SRB-delivering the corresponding write, read, or high-read request. 
For $\textsf{ackW}$, the condition holds because after SRB-delivering the corresponding write, 
the process sets $Lab_i[r,h]=k$ (Line~\ref{NCL:DeliverWrite-Lab}). 
For $\textsf{ackR}$ and $\textsf{hrep}$,
the condition holds because SRB-Support Admissibility of the corresponding read and high-read messages, 
respectively, implies $Lab_i[r,h]=k$.
In all three cases, the set carried by the reply is
$Knw_i[r,k]$, so the condition in Line~\ref{NCL:ValidFunc-5} holds.
\qedhere
\end{enumerate}
\end{proof}

We next prove that if a message is admissible at \emph{some} correct process at time $T$,
then it is admissible at \emph{every} correct process by time $T+2$;
this is used to establish \emph{SRB-Relay (timed)}, in~(\ref{clause:SRB}).
The inductive proof carries with it several additional claims 
about the change of local variables at correct processes. 

\begin{lemma}
\label{lemma:propagation}
The following properties hold for all correct processes $p_i$ and $p_j$, 
processes $p_h$ and $p_{\ell}$, levels $r$, labels $k$,
sets $H$, messages $(P,S,tag)$, and times $T$:

\begin{enumerate}

\item \label{clause:1}
If $\textsc{Admis}(P,S,tag,h)$ holds at $p_i$ by time $T$, 
then it holds at $p_j$ by time $T+2$.

\item \label{clause:SRB}
If $(P,S,tag,h)$ is SRB-delivered at $p_i$ by time $T$, 
then it is SRB-delivered at $p_j$ by time $T+2$.

\item \label{clause:2}
If $H\subseteq Knw_i[r,k]$ by time $T$, then
$H\subseteq Knw_j[r,k]$ by time $T+2$.

\item \label{clause:4}
If $Lab_i[r,h]=k$ by time $T$, then
$Lab_j[r,h]=k$ by time $T+2$.

\item \label{clause:3}
If $Lwr_i[r,k,h]=H$ by time $T$, then
$Lwr_j[r,k,h]=H$ by time $T+2$.

\item \label{clause:5}
If $\mathit{ackW}_i[r,\ell,h]=H$ by time $T$, then
$\mathit{ackW}_j[r,\ell,h]=H$ by time $T+2$.

Similarly, 
if $\mathit{ackR}_i[r,\ell,h]=H$ by time $T$, then
$\mathit{ackR}_j[r,\ell,h]=H$ by time $T+2$.

\end{enumerate}
\end{lemma}

\begin{proof}
We first note that (\ref{clause:SRB}) immediately follows from
(\ref{clause:1}): 
If a message is SRB-delivered by $p_i$ by time $T$,
then it is RB-delivered before time $T$, 
and by RB-Relay (see Lemma~\ref{lemma:RB relay}), 
it is RB-delivered by $p_j$ by time $T+2$.
By $(\ref{clause:1})$,
the message is also admissible at $p_j$ by time $T+2$,
which SRB-delivers it by time $T+2$.

We prove the other claims simultaneously by induction on the level $r$.

\noindent
\textbf{For the base case,} level~$0$, 
we only need to consider the delivery of \textsf{init} messages.
To prove $(\ref{clause:1})$,
if $\textsc{Admis}(P,S,\textsf{init},h)$
holds at $p_i$, then $P=(h,x_h)$, $S=\emptyset$, and $x_h\in X$.
Clearly, these conditions also hold at $p_j$.

To prove $(\ref{clause:2})$, 
suppose that $H\subseteq Knw_i[0,n]$ by time $T$.
Every pair $(h,x_h)\in H$ was added to $Knw_i[0,n]$ upon
SRB-delivery of the corresponding \textsf{init} message from $p_h$.
By (\ref{clause:SRB}), 
$p_j$ SRB-delivers the same \textsf{init} message from $p_h$
and adds $(h,x_h)$ to
$Knw_j[0,n]$ by time $T+2$.
Therefore, $H\subseteq Knw_j[0,n]$ by time $T+2$.
For any label $k\neq n$, $Knw_i[0,k]$ and $Knw_j[0,k]$ remain empty,
so the claim is immediate.

To prove $(\ref{clause:4})$, 
suppose that $Lab_i[0,h]=n$ by time $T$.
$Lab_i[0,h]$ is set when $p_i$ SRB-delivers the \textsf{init} message from $p_h$, 
since by (\ref{clause:SRB}), 
$p_j$ SRB-delivers the same \textsf{init} message from $p_h$
by time $T+2$ and sets $Lab_j[0,h]=n$.
For $k\neq n$, the premise does not hold.

Finally, no nonempty $Lwr$, $\mathit{ackW}$, or $\mathit{ackR}$
entry is created at level~$0$. 
Thus, $(\ref{clause:3})$ and $(\ref{clause:5})$ either have a
false premise or concern entries that retain their common initial
value $\emptyset$, and hence hold trivially.

\noindent
\textbf{For the induction step,} fix a level $r\ge 1$, and assume that all
the claims hold for every level smaller than $r$.

\smallskip
\noindent\emph{Write messages.}
We first prove $(\ref{clause:1})$ for write messages of level $r$.
Suppose that $\textsc{Admis}((H,k,class),S,(\textsf{write},r),h)$ 
holds at $p_i$ by time $T$.
This and Lemma \ref{lemma:RB relay} imply (\ref{clause:SRB}) for write messages.

If $class=\textsf{high}$, let $r'=r-1$ and $k'=k-\frac{f'}{2^r}$.
By Line~\ref{NCL:ValidFunc-3},
$Lab_i[r',h]=k'$ and $H\subseteq Knw_i[r',k']$ by time $T$.
By the induction hypothesis for 
$(\ref{clause:4})$ and $(\ref{clause:2})$, respectively,
$Lab_j[r',h]=k'$ and $H \subseteq Knw_j[r',k']$ by time $T+2$.
The remaining condition $r > 1$ depends only on the message.
Therefore, 
$\textsc{Admis}((H,k,\textsf{high}),S,(\textsf{write},r),h)$
holds at $p_j$ by time $T+2$.

Suppose that $class=\textsf{low}$.
Let $r'=r-1$ and $k'=k+\frac{f'}{2^r}$.
If $r=1$, then $k'=k+\frac{f'}{2}=n$, 
and admissibility requires
$H\subseteq Knw_i[0,n]$ and $|H|\ge n-f$.
By the induction hypothesis for $(\ref{clause:2})$,
$H\subseteq Knw_j[0,n]$ by time $T+2$, 
while the size condition is independent of the local state. 
Hence, $\textsc{Admis}((H,k,\textsf{low}),S,(\textsf{write},r),h)$
holds at $p_j$ by time $T+2$.

If $r>1$, then by the admissibility condition,
$H\subseteq Knw_i[r',k']$,
$H=Lwr_i[r',k',h]$, 
and $\Call{ConfSup}{S,h,r'}$ 
hold by time $T$.
By the induction hypothesis for
$(\ref{clause:2})$ and $(\ref{clause:3})$,
$H\subseteq Knw_j[r',k']$, 
and $H=Lwr_j[r',k',h]$ hold by time $T+2$.

Since $\Call{ConfSup}{S,h,r'}$ holds at $p_i$,
we have that $|S|\ge n-f$ and, 
for every $\langle q,R\rangle\in S$, $R \neq \emptyset$ and
$ackR_i[r',h,q]=R$. 
By the induction hypothesis for $(\ref{clause:5})$,
$ackR_j[r',h,q]=R$ and $R \neq \emptyset$
for every such item by time $T+2$.
Therefore, $\Call{ConfSup}{S,h,r'}$ holds at $p_j$ by time $T+2$.
The remaining conditions depend only on $P$ and $S$, 
so it is admissible at $p_j$ by time $T+2$.

\smallskip
\noindent\emph{Propagation of write information.}
We next use the write case just proved to establish
$(\ref{clause:2})$, $(\ref{clause:4})$, and
$(\ref{clause:3})$ for level $r$.

To prove $(\ref{clause:2})$, suppose that
$H\subseteq Knw_i[r,k]$ by time $T$.
Every pair in $H$ appears in a write message with
label $k$ that $p_i$ SRB-delivers by time $T$.
By the write case of (\ref{clause:SRB}),
$p_j$ SRB-delivers these messages and adds their contents to
$Knw_j[r,k]$ by time $T+2$.
It follows that $H\subseteq Knw_j[r,k]$ by time $T+2$.

To prove $(\ref{clause:4})$, suppose that
$Lab_i[r,h]=k$ by time $T$.
$Lab_i[r,h]=k$ is set when $p_i$ SRB-delivers a write message
$((H,k,class),S,(\textsf{write},r))$ from $p_h$.
By the write case of (\ref{clause:SRB}),
$p_j$ SRB-delivers the message and sets
$Lab_j[r,h]=k$ by time $T+2$.

The proof of $(\ref{clause:3})$ is similar.
If $Lwr_i[r,k,h]=H$ by time $T$, then $p_i$ SRB-delivered a write
message $((H,k,class),S,(\textsf{write},r))$ from $p_h$ by time $T$.
By the write case of $(\ref{clause:SRB})$,
$p_j$ SRB-delivers the same message by time $T+2$ and sets
$Lwr_j[r,k,h]=H$.

\smallskip
\noindent\emph{Reply messages.}
We next prove $(\ref{clause:1})$ for \textsf{ackW}, \textsf{ackR} or \textsf{hrep} messages.
Since $\textsc{Admis}((H,k),\emptyset,(type,r,\ell),h)$ holds at $p_i$ by time $T$, 
Line~\ref{NCL:ValidFunc-5} implies that 
$H\subseteq Knw_i[r,k]$ and $Lab_i[r,\ell]=k$ by time $T$.
By $(\ref{clause:2})$ and $(\ref{clause:4})$ for level~$r$,
$H\subseteq Knw_j[r,k]$ and $Lab_j[r,\ell]=k$ by time $T+2$.
Therefore, $\textsc{Admis}((H,k),\emptyset,(type,r,\ell),h)$ 
holds at $p_j$ by time $T+2$.
This and Lemma \ref{lemma:RB relay} imply (\ref{clause:SRB}) for these messages.

Next, to prove $(\ref{clause:1})$ for \textsf{hreq},
assume $\textsc{Admis}(k,S,(\textsf{hreq},r),h)$ 
holds at $p_i$ by time $T$.
By Line~\ref{ackW},
$S\subseteq Knw_i[r,k]$ and $Lab_i[r,h]=k$ by time $T$.
By $(\ref{clause:2})$ and $(\ref{clause:4})$ for level~$r$,
$S\subseteq Knw_j[r,k]$ and $Lab_j[r,h]=k$ by time $T+2$.
Hence,
$\textsc{Admis}(k,S,(\textsf{hreq},r),h)$ 
holds at $p_j$ by time $T+2$.
This and Lemma \ref{lemma:RB relay} imply (\ref{clause:SRB})
for \textsf{hreq} messages.

\smallskip
\noindent\emph{Propagation of acknowledgments.}
We next establish $(\ref{clause:5})$ for level~$r$ for $\mathit{ackW}$;
the same argument applies for $\mathit{ackR}$.
If $ackW_i[r,\ell,h]=H$ by time $T$, 
then $p_i$ SRB-delivered $((H,k),\emptyset,(ackW,r,\ell))$ 
from $p_h$, for some label $k$.
By (\ref{clause:SRB}) for \textsf{ackW} messages,
$p_j$ SRB-delivers the message by time $T+2$ 
and sets $ackW_j[r,\ell,h]=H$. 

\smallskip
\noindent\emph{Read messages.}
Finally, we prove $(\ref{clause:1})$ for \textsf{read} messages
at level~$r$.
If $\textsc{Admis}(k,S,(\textsf{read},r),h)$ 
holds at $p_i$ by time $T$, 
then $Lab_i[r,h]=k$ and 
$\Call{ConfRead}{S,h,r,k}$ holds at $p_i$ (Line~\ref{NCL:ValidFunc-4}). 
The latter implies that 
$|\{q:S[q]\neq\emptyset\}|\ge n-f$,
$Lwr_i[r,k,h]\neq\emptyset$, and, 
for every $q$ with $S[q]\neq\emptyset$, $S[q]=ackW_i[r,h,q]$.
By $(\ref{clause:3})$,
$Lwr_j[r,k,h]\neq\emptyset$ by time $T+2$.
By $(\ref{clause:5})$, 
for every $q$ with $S[q]\neq\emptyset$,
$S[q]=ackW_j[r,h,q]$ by time $T+2$.
The cardinality condition depends only on the message itself, 
and hence, 
$\Call{ConfRead}{S,h,r,k}$ holds at $p_j$ by time $T+2$.
By $(\ref{clause:4})$,
$Lab_j[r,h]=k$ by time $T+2$,
so $\textsc{Admis}(k,S,(\textsf{read},r),h)$ 
holds at $p_j$ by time $T+2$.

This completes the induction step and the proof.
\end{proof}

\begin{restatable}{lemma}{SRBvalid}
\label{lemma:SRB validity}
Algorithm~\ref{alg:SRB}, with \textsc{Admis} of Algorithm~\ref{alg:clean-admis}, 
satisfies SRB-Validity (timed).
\end{restatable}

\begin{proof}
By Lemma~\ref{lemma:RB validity} every correct process RB-delivers the message by time $T+3$. 
By Lemma~\ref{lemma:local-admissibility} the message is admissible at the sender at time $T$, 
so by Lemma~\ref{lemma:propagation}(\ref{clause:1}), 
it is admissible at every correct process by time $T+2$,
which SRB-delivers the message by time $T+3$.
\end{proof}

\subsection{Classifier properties}

We start by stating the properties of the classifier procedures 
that are executed in each level with a given label.
For this proof, we fix a label $k$ at level $r$.
For every level $r$, $1\leq r<F$, 
let $G(r,k)$ denote the set of processes whose write message 
for level $r$ carries label $k$.
For convenience, $G(F,k)$ is the set of processes whose label after completing level $F-1$ is $k$.
Each process in $G(r,k)$ is classified either $low$ or $high$.
The former set is denoted $L(r,k)$, 
and it is equal to $G(r+1,k-\frac{f'}{2^{r+1}})$; 
its label at level $r+1$ is $l(r,k)=k-\frac{f'}{2^{r+1}}$.
The latter set is denoted $H(r,k)$,
and it is equal to $G(r+1,k+\frac{f'}{2^{r+1}})$; 
its label at level $r+1$ is $h(r,k)=k+\frac{f'}{2^{r+1}}$.
(See the right-hand side of Figure~\ref{figure:classifiers tree}.)

The proof makes use of a simple \emph{quorum intersection property} 
for two sets $Q_1$ and $Q_2$: 
\begin{quote}
    If $n > 3f$, $|Q_1| \geq n-f$ and $|Q_2| \geq n-f$, 
    then $| Q_1 \cap Q_2 | > f$ and 
    there is a correct process in $Q_1 \cap Q_2$.
\end{quote}

We assume process $p_i$ starts level $r \geq 1$ with a set $V_{i,r}$;
for $r=0$, define $V_{i,0} = \{(i,x_i)\}$.
The following lemma shows that these sets do not decrease across levels.
It clearly holds for a low process that keeps its current set.
For a high process, it follows since the quorum acknowledging its write intersects the quorum responding to its high-read request.

\begin{lemma}\label{lemma:inclusion}
For every correct process $p_i$ and every level $r \geq 0$, 
$V_{i,r} \subseteq V_{i,r+1}$.
\end{lemma}

\begin{proof} 
For $r=0$, the claim holds by the condition in Line~\ref{MA:Init-1}.

For $r > 0$, note that a correct low process keeps its input unchanged 
(Line~\ref{NCL:Classify-5}).

A correct high process $p_i$ SRB-broadcasts a write message carrying $V_{i,r}$ (Line~\ref{NCL:WriteStep-4}). 
Let $Q_1$ be the set of $n-f$ processes from which $p_i$ SRB-delivers an \textsf{ackW} message. 
Before sending an \textsf{ackW} message (Line~\ref{NCL:DeliverWrite-4}), 
a correct process $p_j \in Q_1$ adds $V_{i,r}$ into its $\textit{Knw}_j[r,k]$ (Line~\ref{NCL:DeliverWrite-2}). 
After receiving \textsf{ackW} messages from $Q_1$, 
$p_i$ SRB-broadcasts an hreq message.
Let $Q_2$ be the set of $n-f$ processes from which $p_i$ 
SRB-delivers an \text{hrep} message.
By the quorum intersection property, 
there is a correct process $p_h \in Q_1 \cap Q_2$.

Then $p_h$ received $p_i$'s write message before sending 
an \textsf{hrep} message to $p_i$.
By Invariant~\ref{invariant:know monotonicity}, 
$V_{i,r}$ is contained in the set $\textit{Knw}_h[r,k]$
that $p_h$ sends to $p_i$ (Line~\ref{NCL:DeliverHigh-3}),
and the lemma follows since $V_{i,r+1}$ is the union of these sets. 
\end{proof}

For label $k$ at level $r$, $0 < r < F$, 
$DV_k^r$ is the set of pairs $(p,v) \in V$ such that 
a message $((V,k,class), S, (\text{write},r))$
is SRB-delivered by some correct process;
note that this process is in $G(r,k)$. 
For a label $k$ at the final level $F$, 
$DV_k^F = \bigcup\{ V_{i,F} : \text{$p_i$ is a correct process in $G(F,k)$}\}$.

The next lemma shows that a pair is delivered in a classifier,
only if it was delivered in its parent classifier. 
For internal levels, the lemma follows directly from admissibility of the child write; the final level uses the definition of $DV_F$. 

\begin{lemma}
\label{lemma:properties p3 and p4}
If $k$ is a label at level $r$, $1\le r<F$, 
and $x$ is either $l(r,k)$ or $h(r,k)$,
then $DV_{x}^{r+1} \subseteq DV_{k}^{r}$.
\end{lemma}
\begin{proof} 
First, suppose $r < F-1$.
Consider a pair $(p,v) \in DV_{x}^{r+1}$.
Then a message $((V,x,class),S,(\text{write},r+1))$ with $(p,v) \in V$ 
is SRB-delivered by some correct process $p_j$ at level $r+1$.
By Proposition~\ref{prop:SRB support admissibility}, 
$\textsc{Admis}((V,x,class),S,(\text{write},r+1))$ holds at process $p_j$.

If $\textit{class} = \textit{low}$, define $k'= l(r,k)+f'/2^{r+1}$,
and if $\textit{class} = \textit{high}$, define $k' = h(r,k)-f'/2^{r+1}$.
In either case, the corresponding admissibility condition
(first clause of Line~\ref{NCL:ValidFunc-2} or 
second clause in Line~\ref{NCL:ValidFunc-3})
implies that $V\subseteq Knw_j[r,k']$.
By Invariant~\ref{invariant:know monotonicity},
every pair in $Knw_j[r,k']$ originates from a write message
SRB-delivered by $p_j$ with label $k'$.
Hence, $(p,v)$ appears in a write message
SRB-delivered by the correct process $p_j$
at level $r$ with label $k'$,
implying that $(p,v) \in DV_{k'}^r$.
By the definitions of $l(r,k)$ and $h(r,k)$,
$k'=k$ in either case, and $(p,v) \in DV_k^r$.

Now suppose that $r=F-1$, and let $(p,v)\in DV_x^F$.
By definition of $DV_x^F$, there is a correct process
$p_i\in G(F,x)$ such that $(p,v)\in V_{i,F}$.

If $p_i$ is classified low at level $F-1$, then
$V_{i,F}=V_{i,F-1}$.
Since $p_i$ is correct, its level-$F-1$ write carrying $V_{i,F-1}$
is SRB-delivered by every correct process
(Lemma~\ref{lemma:SRB validity}).
Hence, $V_{i,F}\subseteq DV_k^{F-1}$.

If $p_i$ is classified high at level $F-1$, then $V_{i,F}$ is the
union of the sets carried by the \textsf{hrep} messages it
SRB-delivers.
By SRB-Support Admissibility
(Proposition~\ref{prop:SRB support admissibility}) and the
admissibility condition for \textsf{hrep} messages, each such set
is contained in $Knw_i[F-1,k]$.
Every pair in $Knw_i[F-1,k]$ originates from a level-$F-1$ write with
label $k$ that is SRB-delivered by the correct process $p_i$.
Thus, $V_{i,F}\subseteq DV_k^{F-1}$. 

In either case, $(p,v)\in DV_k^r$, proving the lemma.
\end{proof}

\begin{lemma}\label{lemma:bounds}
Suppose $k$ is a label at level $r$, $1 \leq r < F$, 
and $L$ and $H$ are two nonnegative integers such that:
(a) $L\leq k\leq H$,
(b) $L<\lvert V_{i,r} \rvert \leq H$ for every correct process $p_i$, and
(c) $\lvert DV_k^r \rvert\leq H$.
Then the following properties hold:
\begin{enumerate}
\item  $\lvert DV_{h(r,k)}^{r+1} \rvert \leq H$. \label{p6}
\item  $\lvert DV_{l(r,k)}^{r+1} \rvert \leq k$. \label{p5} 
\item $\lvert\bigcup \{  V_{i,r+1}  : p_i \in H(r,k)$ and $p_i$ is correct $\}\rvert \leq H$. \label{p10}
\item $\lvert \bigcup \{  V_{i,r+1}  : p_i \in L(r,k)$ and $p_i$ is correct $\}\rvert \leq k$. \label{p11}
\item for every correct process $p_i \in H(r,k)$, $k<\lvert V_{i,r+1} \rvert \leq H$. \label{p1}
\item for every correct process $p_i \in L(r,k)$, 
$L < \lvert V_{i,r+1} \rvert \leq k$. \label{p2} 
\end{enumerate}
\end{lemma}

\begin{proof}
\begin{enumerate}

\item 
By Lemma~\ref{lemma:properties p3 and p4}, 
$DV_{h(r,k)}^{r+1} \subseteq DV_k^r$;\
by (c), 
$\lvert DV_k^r \rvert \leq H$, implying (\ref{p6}).

\item
First suppose that $r<F-1$.
Let $P$ be the set of processes that SRB-broadcast a message
$((V,l(r,k),\textit{low}),S,(\text{write},r+1))$
that is SRB-delivered by some correct process.
Thus, $DV_{l(r,k)}^{r+1}=\bigcup_{p_h\in P}V_{h,r+1}$.

The level-$(r+1)$ write of every process $p_h$ is admissible
at some correct process.
By SRB-Support Admissibility,
the condition in
Line~\ref{NCL:ValidFunc-2} holds.
In particular, its fifth clause implies that a level-$r$ write by $p_h$,
carrying the same set, was SRB-delivered.
Hence, $V_{h,r+1}=V_{h,r}$, even when $p_h$ is not correct.

The second admissibility clause implies that
the support of this write contains at least $n-f$ non empty
\textsf{ackR} messages addressed to $p_h$.
Consider one such message sent by a correct process $p_\ell$.
The admissibility of the corresponding read implies that its support
contains at least $n-f$ \textsf{ackW} messages.
Let $Q_h$ be the set of processes that sent these \textsf{ackW} messages.
Every correct process $p_s\in Q_h$ adds $V_{h,r}$ to
$Knw_s[r,k]$ before broadcasting its \textsf{ackW};
hence, $V_{h,r}\subseteq Knw_s[r,k]$ thereafter.

For each process $p_a$ in $P$, consider the first event at which 
a correct process SRB-broadcasts an \textsf{ackR} message contained 
in the support of $p_a$'s level-$(r+1)$ write.
Such an event exists, since this support contains at least $n-f>f$
\textsf{ackR} messages.
Let $p_j$ be such that this event is latest in the execution order,
and let $Q'$ be the set of senders of the \textsf{ackR} messages
contained in the support of $p_j$'s level-$(r+1)$ write.

By the quorum-intersection property, $Q_h \cap Q'$ contains a correct process $p_s$.
The \textsf{ackW} message from $p_s$ is contained in the support of $p_h$'s read.
Hence it is SRB-broadcast before the first correct \textsf{ackR}
message appearing in the support of $p_h$'s next-level write.
By the choice of $p_j$, this event occurs no later than the first correct
\textsf{ackR} event in the support of $p_j$'s write, which in turn occurs
no later than the \textsf{ackR} message from $p_s$ contained in that support.
Thus, $p_s$ SRB-broadcasts its \textsf{ackW} message before its \textsf{ackR} message to $p_j$.

This holds for every $p_h\in P$.
Therefore, the union of the sets carried by the \textsf{ackR} messages
contained in the support of $p_j$'s write contains 
$\bigcup_{p_h\in P}V_{h,r}
=\bigcup_{p_h\in P}V_{h,r+1}$.
By the fourth clause of Line~\ref{NCL:ValidFunc-2}, the union of the
sets carried by these \textsf{ackR} messages has size at most $k$.
Thus, $\left|\bigcup_{p_h\in P}V_{h,r+1}\right|\leq k$.
Since $DV_{l(r,k)}^{r+1}=\bigcup_{p_h\in P}V_{h,r+1}$,
it follows that 
$|DV_{l(r,k)}^{r+1}|\leq k$.

Assume $r=F-1$. By definition,
$DV^F_{l(r,k)} = \bigcup\{V_{h,F}: \text{$p_h$ is a correct process in $L(r,k)$}\}$. 
If there is no correct process in $L(r,k)$, the claim is immediate.
Otherwise, let $p_j$ be the last correct process in $L(r,k)$ to
SRB-broadcast its level-$r$ \textsf{read}.

For every correct $p_h\in L(r,k)$, the same quorum-intersection
argument used above shows that $V_{h,r}\subseteq ReadValues_j$: 
$p_h$'s read is supported by at least $n-f$ \textsf{ackW} messages,
$p_j$ receives at least $n-f$ \textsf{ackR} messages, and
$p_h$ broadcasts its read no later than $p_j$.
Since every correct low process keeps its input set,
$V_{h,F}=V_{h,r}$, and since $p_j$ is classified low,
$|ReadValues_j|\le k$.
Therefore, $|DV^F_{l(r,k)}|\le k$.

\item
For $r = F-1$, the claim follows from (\ref{p6}) 
and the definition of $DV^F$; so, assume $r < F-1$.
We show that $V_{i,r+1}\subseteq DV_{h(r,k)}^{r+1}$
for every correct process $p_i\in H(r,k)$.
Consider a pair $(p,v)\in V_{i,r+1}$.
Since $p_i$ is classified as high at level $r$, 
it SRB-broadcasts the message
$((V_{i,r+1},h(r,k),\textit{high}),Sup_i[r],(\textsf{write},r+1))$
(Line~\ref{NCL:WriteStep-4}).
By SRB-Validity (Lemma~\ref{lemma:SRB validity}), 
this message is SRB-delivered by every correct process.
Hence, $(p,v)\in DV_{h(r,k)}^{r+1}$.

Therefore,
$\bigcup\{V_{i,r+1}:p_i\in H(r,k)\text{ and }p_i\text{ is correct}\}
\subseteq DV_{h(r,k)}^{r+1}$.
By (\ref{p6}), $|DV_{h(r,k)}^{r+1}|\leq H$.

\item

For $r = F-1$, the claim follows from (\ref{p5}) 
and the definition of $DV^F$; so, assume $r < F-1$.
Consider a pair $(p,v)\in V_{i,r+1}$.
Since $p_i$ is classified as low at level $r$,
$V_{i,r+1}=V_{i,r}$ by Line~\ref{NCL:Classify-5}.
Process $p_i$ therefore SRB-broadcasts the message
$((V_{i,r+1},l(r,k),\textit{low}),Sup_i[r],(\textsf{write},r+1))$
(Line~\ref{NCL:WriteStep-4}).
By SRB-Validity (Lemma~\ref{lemma:SRB validity}), this message is
SRB-delivered by every correct process.
Hence, $(p,v)\in DV_{l(r,k)}^{r+1}$.

Therefore,
$\bigcup\{V_{i,r+1}:p_i\in L(r,k)\text{ and }p_i\text{ is correct}\}
\subseteq DV_{l(r,k)}^{r+1}$.
By (\ref{p5}), $|DV_{l(r,k)}^{r+1}|\leq k$.

\item 
By (\ref{p10}), $|V_{i,r+1}|\leq H$.

To show that $|V_{i,r+1}|>k$, recall that a high process $p_i$
sets $V_{i,r+1}$ to the union of the $n-f$ sets carried by the
\textsf{hrep} messages it SRB-delivers
(Line~\ref{NCL:Classify-4}).
Since $n-f>f$, at least one of these messages is sent by a correct
process $p_j$.

Process $p_j$ SRB-broadcasts its \textsf{hrep} message to $p_i$
after SRB-delivering the \textsf{hreq} message from $p_i$.
This \textsf{hreq} message carries $\textit{ReadValues}_i$
(Line~\ref{NCL:Classify-2}).
By SRB-Support Admissibility
and Lines~\ref{ackW} and~\ref{NCL:DeliverHigh-3},
the set carried by $p_j$'s \textsf{hrep} message contains
$\textit{ReadValues}_i$.
Since $|\textit{ReadValues}_i|>k$, it follows that
$|V_{i,r+1}|>k$.

\item 
Since $p_i\in L(r,k)$ is correct,
$V_{i,r+1}=V_{i,r}$ by Line~\ref{NCL:Classify-5}.
By (b), 
$L<|V_{i,r}|$,
so it remains to show that $|V_{i,r}|\le k$.

In level $r$, $p_i$ first SRB-broadcasts a \textsf{write} message
carrying $V_{i,r}$ (Line~\ref{NCL:WriteStep-4}).
Let $Q_w$ be a set of $n-f$ processes from which $p_i$
SRB-delivers \textsf{ackW} messages
(Line~\ref{NCL:WriteStep-5}).
Then, $p_i$ SRB-broadcasts a \textsf{read} message
(Line~\ref{NCL:1collect-1}).
Let $Q_r$ be a set of $n-f$ processes from which $p_i$
SRB-delivers \textsf{ackR} messages
(Line~\ref{NCL:1collect-2}).
By the quorum intersection property,
there is a correct process $p_j\in Q_r\cap Q_w$.

By Line~\ref{NCL:DeliverWrite-2},
$V_{i,r}\subseteq Knw_j[r,k]$
before $p_j$ SRB-broadcasts its \textsf{ackW} message to $p_i$.
Since $p_j$ SRB-delivers $p_i$'s \textsf{read} message only after
broadcasting this \textsf{ackW} message,
it SRB-broadcasts the corresponding \textsf{ackR} message afterward.
By Line~\ref{NCL:DeliverRead-4},
this \textsf{ackR} message contains $Knw_j[r,k]$.
By Invariant~\ref{invariant:know monotonicity},
$Knw_j[r,k]$ does not decrease.
Hence, $V_{i,r}$ is contained in the \textsf{ackR} message
that $p_i$ SRB-delivers from $p_j$.

Therefore, $V_{i,r}\subseteq \textit{ReadValues}_i$.
Since $p_i$ is classified as low,
$|\textit{ReadValues}_i|\le k$.
It follows that
$|V_{i,r+1}|=|V_{i,r}|\le k$.
\qedhere
\end{enumerate}
\end{proof}

The next lemma shows how the size of $V_{i,r}$ becomes increasingly restricted as the level increases;
its proof uses properties (\ref{p6}) and (\ref{p1}) of Lemma~\ref{lemma:bounds} for high processes, 
and properties (\ref{p2}) and (\ref{p5}) for low processes. 
Recall that $G(r,k)$ are the processes at level $r$ with label $k$. 

\begin{lemma}
\label{lemma:3}
For every level $r$, $1 \leq r < F$ and every correct process 
$p_i\in G(r,k)$,
$k-\frac{f'}{2^r}-1 <\lvert V_{i,r}\rvert\leq k+\frac{f'}{2^r}$,
and $\lvert DV_k^r\rvert\leq k+\frac{f'}{2^r}$.
\end{lemma}

\begin{proof}
By induction on $r$.
In the base case, $r=1$, the only label is $k=n-\frac{f'}{2}$.
Before level~1, each correct process SRB-delivers at least 
$n-f \geq n-f' = k-\frac{f'}{2}$
different values and at most $n = k + \frac{f'}{2}$ values.
Thus,
$k-\frac{f'}{2}-1 < \lvert V_{i,1} \lvert \leq k + \frac{f'}{2}$.
By SRB Integrity and Agreement, at most $n$ distinct values are delivered at correct processes, and therefore
$\lvert DV_{n-\frac{f'}{2}}^1 \rvert \leq n = k + \frac{f'}{2}$.

For the induction step,
consider an arbitrary group $G(r,k)$ at level $r > 1$ with label $k$.
Let $G'(r-1,k')$ be the parent group of $G(r,k)$ at level $r-1$
with label $k'$.
By the induction hypothesis,
for every correct process $p_i\in G'(r-1,k')$,
$k'-\frac{f'}{2^{r-1}}-1<\lvert V_{i,r-1}\rvert
\leq k'+\frac{f'}{2^{r-1}}$,
and $\lvert DV_{k'}^{r-1}\rvert
\leq k'+\frac{f'}{2^{r-1}}$.
Therefore, the conditions of Lemma~\ref{lemma:bounds} hold
for $L=k'-\frac{f'}{2^{r-1}}-1$
and $H=k'+\frac{f'}{2^{r-1}}$.

If $G(r,k)=H(r-1,k')$, then
$k=h(r-1,k')=k'+\frac{f'}{2^r}$.
By Lemma~\ref{lemma:bounds}(\ref{p1}),
for every correct process $p_i\in G(r,k)$,
$k'<\lvert V_{i,r}\rvert\leq H$.
Since $k-\frac{f'}{2^r}-1=k'-1<k'$
and $H=k'+\frac{f'}{2^{r-1}}=k+\frac{f'}{2^r}$,
we have
$k-\frac{f'}{2^r}-1<\lvert V_{i,r}\rvert
\leq k+\frac{f'}{2^r}$.
Furthermore, by Lemma~\ref{lemma:bounds}(\ref{p6}),
$\lvert DV_k^r\rvert\leq H=k+\frac{f'}{2^r}$.

In a similar manner, if $G(r,k)=L(r-1,k')$, then
$k=l(r-1,k')=k'-\frac{f'}{2^r}$.
By Lemma~\ref{lemma:bounds}(\ref{p2}),
for every correct process $p_i\in G(r,k)$,
$L<\lvert V_{i,r}\rvert\leq k'$.
Since
$L=k'-\frac{f'}{2^{r-1}}-1
=k-\frac{f'}{2^r}-1$,
we have
$k-\frac{f'}{2^r}-1<\lvert V_{i,r}\rvert
\leq k'= k+\frac{f'}{2^r}$.
Finally, by Lemma~\ref{lemma:bounds}(\ref{p5}),
$\lvert DV_k^r\rvert\leq k'= k+\frac{f'}{2^r}$.
\end{proof}
\begin{lemma}\label{lemma:low-in-high}
If $k$ is a label at level $r$, $1\le r<F$, 
and $p_i$ is a correct process in $H(r,k)$, then
$V_{j,F}\subseteq V_{i,F}$ for every correct process $p_j$
whose final group is a descendant of $L(r,k)$.
\end{lemma}

\begin{proof}
We first show that $DV^{r+1}_{l(r,k)} \subseteq V_{i,r+1}$. 

Suppose first that $r<F-1$.
A pair $(p,v)$ is in $DV^{r+1}_{l(r,k)}$ if there is a message
$m=((V,l(r,k),\textit{low}),S,(\textsf{write},r+1))$
from process $p_a$ that is
SRB-delivered by a correct process $p_h$, such that $(p,v)\in V$.
By SRB-Support Admissibility,
the condition in Line~\ref{NCL:ValidFunc-2} holds at $p_h$ for $m$.
In particular, its third and fourth clauses imply that
$|V|\leq k$.

Let $Q_\ell$ be the set of processes whose \textsf{ackR} messages
appear in $S$. Then $|Q_\ell|\geq n-f$, and for every
$p_q\in Q_\ell$, $S$ contains an item $\langle q,R_q\rangle$,
where $R_q$ is the set carried by the corresponding
\textsf{ackR} message from $p_q$.

Let $Q_h$ be the set of processes that SRB-broadcast the
\textsf{hrep} messages used by $p_i$ to construct $V_{i,r+1}$
at the end of level $r$ (Line~\ref{NCL:Classify-3});
thus, $|Q_h|\geq n-f$.
By the quorum intersection property, there is a correct process
$p_s\in Q_\ell\cap Q_h$.

Consider the \textsf{hrep} message broadcast by $p_s$ and used by
$p_i$.
Assume first that $p_s$ SRB-broadcasts this
\textsf{hrep} message \emph{before} its \textsf{ackR} message to
$p_a$.
By Line~\ref{NCL:DeliverHigh-3}, $p_s$ broadcasts the \textsf{hrep} message only after SRB-delivering the corresponding \textsf{hreq} message from $p_i$.
By SRB-Support Admissibility
and the
admissibility condition for \textsf{hreq} messages in
Line~\ref{ackW},
$\textit{ReadValues}_i\subseteq Knw_s[r,k]$.
Since $p_i$ broadcasts its \textsf{hreq} message only when
$|\textit{ReadValues}_i|>k$, it follows that
$|Knw_s[r,k]|>k$ at that time.
When $p_s$ later broadcasts its \textsf{ackR} message to $p_a$,
the message carries $Knw_s[r,k]$.
By Invariant~\ref{invariant:know monotonicity}, this set does not
decrease, and hence $|R_s|>k$.
Since $\langle s,R_s\rangle\in S$, we have 
$\left|\bigcup\{R:\langle -,R\rangle\in S\}\right| \geq |R_s|>k$,
contradicting the admissibility condition for $m$ at $p_h$.

It follows that $p_s$ broadcasts its \textsf{ackR} message to $p_a$
before its \textsf{hrep} message to $p_i$.
Before broadcasting this \textsf{ackR} message, $p_s$ SRB-delivers
the corresponding level-$r$ \textsf{read} message from $p_a$.
By SRB-Support Admissibility
and the admissibility condition for this \textsf{read}
(Line~\ref{confRead-check-write}),
$Lwr_s[r,k,a]\neq\emptyset$.
Thus, $p_s$ previously SRB-delivered the level-$r$
\textsf{write} message from $p_a$.
Let $V'$ be the set carried by this write message.
When $p_s$ SRB-delivers it, Line~\ref{NCL:DeliverWrite-2}
adds $V'$ to $Knw_s[r,k]$.
Since $Knw_s[r,k]$ does not decrease, the later \textsf{hrep}
message broadcast by $p_s$ contains $V'$.

On the other hand, by SRB-Support Admissibility for $m$, the
last condition in Line~\ref{NCL:ValidFunc-2} implies that the
level-$r$ \textsf{write} from $p_a$ recorded at $p_h$ carries $V$.
By SRB-Agreement, this is the same level-$r$ \textsf{write} from
$p_a$ that $p_s$ SRB-delivered, and therefore $V'=V$.
Thus, $(p,v)$ appears in the \textsf{hrep} message from $p_s$.
Since this message is one of those whose union forms $V_{i,r+1}$,
we have $(p,v)\in V_{i,r+1}$.

Now suppose that $r=F-1$.
Recall that $f'/2^r=1/2$.
Let $L=k-\frac32$ and $H=k+\frac12$; these are integers.
By Lemma~\ref{lemma:3}, $|DV_k^{F-1}|\leq k+\frac12$,
and the hypotheses of Lemma~\ref{lemma:bounds} hold for
these values of $L$ and $H$.
Thus, Lemma~\ref{lemma:bounds}(\ref{p1}) implies
$k<|V_{i,F}|\leq k+\frac12$. 
Since $k$ is a half-integer, $k+1/2$ is an integer, and hence
$|V_{i,F}|=k+\frac12$.
By the definition of $DV^F$ and Lemma~\ref{lemma:properties p3 and p4},
$V_{i,F}\subseteq DV^F_{h(r,k)}\subseteq DV_k^r$.
Since $V_{i,F}\subseteq DV^r_k$, while
$|V_{i,F}|=k+\frac12$ and $|DV^r_k|\le k+\frac12$,
it follows that $V_{i,F}=DV^r_k$.
Another application of Lemma~\ref{lemma:properties p3 and p4} gives
$DV^F_{l(r,k)}\subseteq DV_k^r=V_{i,F}$.
Thus, in either case, $DV^{r+1}_{l(r,k)}\subseteq V_{i,r+1}$.

Finally, let $p_j$ be a correct process whose final group is a
descendant of $L(r,k)$, and let $k_j$ be its label at level $F$.
By definition, $V_{j,F}\subseteq DV^F_{k_j}$. 
Repeated application of Lemma~\ref{lemma:properties p3 and p4} along the path from $L(r,k)$ to the final group of $p_j$ gives
$DV^F_{k_j}\subseteq DV^{r+1}_{l(r,k)}$.

By the first part of the proof and Lemma~\ref{lemma:inclusion},
$V_{j,F}
    \subseteq DV^F_{k_j}
    \subseteq DV^{r+1}_{l(r,k)}
    \subseteq V_{i,r+1}
    \subseteq V_{i,F}$.
\end{proof}

We complete the analysis of the classifier by observing that a correct
process that starts a classifier at time $T$ completes it by time
$T+18$:
There are at most three request-response phases:
\textsf{write}-\textsf{ackW}, \textsf{read}-\textsf{ackR}, and,
for a high process, \textsf{hreq}--\textsf{hrep}.
Each phase consists of two consecutive SRB-broadcasts and therefore
takes at most $2\cdot 3$ time units, by
Lemma~\ref{lemma:SRB validity}.

\subsection{Putting the pieces together}

Recall that $f'$ is the smallest power of 2 greater than or equal to $f$ and that $F= \log_2 f' +2$. 

\begin{lemma}\label{lemma:same-leaf}
For any two correct processes $p_i$ and $p_j$ in the same group $G(F,k)$, $V_{i,F} = V_{j,F}$.
\end{lemma}

\begin{proof} 
Let $G'=G(F-1,k')$ be the parent of $G(F,k)$.
Since $F-1=\log_2 f'+1$, we have
$f'/2^{F-1}=1/2$, and $k'$ is a half-integer.
Let $L=k'-3/2$ and $H=k'+1/2$; both $L$ and $H$ are integers.

By Lemma~\ref{lemma:3}, for every correct process
$p_h\in G'$,
$L<|V_{h,F-1}|\leq H$, and
$|DV_{k'}^{F-1}|\leq H$.
Thus, the hypotheses of Lemma~\ref{lemma:bounds} hold.

Suppose first that $G(F,k)=H(F-1,k')$.
By Lemma~\ref{lemma:bounds}(\ref{p1}), every correct process
$p_h$ in this group satisfies
$k'<|V_{h,F}|\leq H$.
Since $k'$ is a half-integer and $H=k'+1/2$ is an integer,
it follows that $|V_{h,F}|=H$.
By Lemma~\ref{lemma:bounds}(\ref{p10}), the union of the sets
of all correct processes in this group has size at most $H$.
Hence, all these sets are equal.

Suppose next that $G(F,k)=L(F-1,k')$.
By Lemma~\ref{lemma:bounds}(\ref{p2}), every correct process
$p_h$ in this group satisfies
$L<|V_{h,F}|\leq k'$.
Since $L$ is an integer and $k'=L+3/2$,
it follows that $|V_{h,F}|=L+1$.
By Lemma~\ref{lemma:bounds}(\ref{p11}), the union of the sets
of all correct processes in this group has size at most $k'$.
Since its size is integral, it is at most
$\lfloor k'\rfloor=L+1$.
Hence, all these sets are equal.
\end{proof}

\begin{lemma}[Comparability]\label{lemma:comparability}
For every pair of correct processes $p_i$ and $p_j$, 
the sets $V_{i,F}$ and $V_{j,F}$ are comparable.
\end{lemma}

\begin{proof}
If $p_i$ and $p_j$ belong to the same final group, then
Lemma~\ref{lemma:same-leaf} implies that $V_{i,F}=V_{j,F}$.
Otherwise, let $(r,k)$ be their last common classifier,
so that $p_i\in L(r,k)$ and $p_j\in H(r,k)$.
The final group of $p_i$ is a descendant of $L(r,k)$, 
and Lemma~\ref{lemma:low-in-high} implies that
$V_{i,F}\subseteq V_{j,F}$.
\end{proof}

\begin{restatable}[Byzantine Upward-Validity]{lemma}{ByzUpwardValid}
\label{lemma:byz upward validity}
Let $C$ be the set of correct processes.
There exists a set $B$ of processes, with $B \cap C=\emptyset$
and $|B|\leq f$, and a value $v_j\in X$ for every $p_j\in B$,
such that
$\bigcup_{p_i\in C} V_{i,F}
\subseteq
\left(\bigcup_{p_i\in C}\{(i,x_i)\}\right)
\cup
\left(\bigcup_{p_j\in B}\{(j,v_j)\}\right)$. 
\end{restatable}
\begin{proof}
Let $B$ be the set of Byzantine processes from which some correct
process SRB-delivers an \textsf{init} message.
By SRB-Agreement (Lemma~\ref{lemma:SRB agreement}),
all correct processes that SRB-deliver an \textsf{init} message
from a process $p_j\in B$ deliver the same value; 
denote this value by $v_j$.
Since there are at most $f$ Byzantine processes, $|B|\leq f$.
Define $\mathcal{U}\triangleq
\left(\bigcup_{p_i\in C}\{(i,x_i)\}\right)
\cup
\left(\bigcup_{p_j\in B}\{(j,v_j)\}\right)$.

We show by induction on $r$ that
$DV_k^r\subseteq\mathcal{U}$ for every label $k$ at level $r$.
In the base case,  
$DV_{n-\frac{f'}{2}}^1\subseteq\mathcal{U}$,
since every pair appearing in a level~1 write originates from an
SRB-delivered \textsf{init} message.
For the induction step, let $k$ be a label at level $r+1$, and let
$k'$ be the label of its parent at level $r$.
By Lemma~\ref{lemma:properties p3 and p4},
$DV_k^{r+1}\subseteq DV_{k'}^r$.
The induction hypothesis therefore implies that
$DV_k^{r+1}\subseteq\mathcal{U}$.

Finally, let $p_i$ be a correct process, and let $k$ be its label at
level $F$. By definition, $V_{i,F}\subseteq DV_k^F$.
Since $DV_k^F\subseteq U$, we have $V_{i,F}\subseteq U$.
Taking the union over all correct processes proves the lemma.
\end{proof}

A correct process returns $\sqcup \{ x \mid (j,x) \in V_{i,F} \}$.
\emph{Downward-validity} follows from Line~\ref{MA:Init-1},
and inductive application of Lemma~\ref{lemma:inclusion}.
Lemma~\ref{lemma:comparability} and Lemma~\ref{lemma:byz upward validity} imply \emph{comparability} and {Byzantine upward-validity}, respectively. 

A correct process $p_i$ starts by sending an \textsf{init} message, 
which by Lemma~\ref{lemma:SRB validity} is SRB-delivered at all correct processes within 3 time units; 
each such delivery adds $p_i$'s pair to $Knw_i$. 
Thus, within 3 time units after the last correct process starts, 
the condition in Line~\ref{MA:Init-1} holds for all correct processes. 
A process then goes through $F-1 \in O(\log f)$ classifiers,
each requiring $O(1)$ time, which implies:

\begin{theorem}
There is an algorithm that solves Byzantine lattice agreement, 
for $f$ failures and $n > 3f$ processes.
The algorithm takes $O( \log f )$ time, and does not use authentication.
\end{theorem}

%% file: references.bib
@article{AttiyaBD1995,
  title = {Sharing memory robustly in message-passing systems},
  author = {Attiya, Hagit and Bar-Noy, Amotz and Dolev, Danny},
  journal = {Journal of the ACM (JACM)},
  volume = {42},
  number = {1},
  pages = {124--142},
  year = {1995},
  publisher = {ACM New York, NY, USA},
  timestamp = {Mon, 16 Sep 2019 01:00:00 +0200},
  biburl = {https://dblp.org/rec/journals/jacm/AttiyaBD95.bib},
  bibsource = {dblp computer science bibliography, https://dblp.org},
  doi = {10.1145/200836.200869},
  _bib2doi_selected = {dblp:/rec/journals/jacm/AttiyaBD95.bib},
  _bib2doi_confirmed = {true},
}

@article{Attiya_atomicHR1995,
  title = {Atomic Snapshots Using Lattice Agreement},
  volume = {8},
  issn = {1432-0452},
  url = {https://doi.org/10.1007/BF02242714},
  doi = {10.1007/BF02242714},
  language = {en},
  number = {3},
  urldate = {2024-03-04},
  journal = {Distributed Comput.},
  author = {Hagit Attiya and Maurice Herlihy and Ophir Rachman},
  month = mar,
  year = {1995},
  keywords = {Atomic snapshot, Lattice agreement},
  pages = {121--132},
  timestamp = {Fri, 13 Mar 2020 00:00:00 +0100},
  biburl = {https://dblp.org/rec/journals/dc/AttiyaHR95.bib},
  bibsource = {dblp computer science bibliography, https://dblp.org},
  _bib2doi_selected = {dblp:/rec/journals/dc/AttiyaHR95.bib},
  _bib2doi_confirmed = {true},
  _bib2doi_finished = {true},
}

@article{attiya_atomic_1998,
  title = {Atomic Snapshots in {$O(n \log n)$} Operations},
  volume = {27},
  issn = {0097-5397},
  doi = {10.1137/S0097539795279463},
  number = {2},
  urldate = {2024-03-14},
  journal = {SIAM J. Comput.},
  author = {Attiya, Hagit and Rachman, Ophir},
  month = jan,
  year = {1998},
  publisher = {Society for Industrial and Applied Mathematics},
  pages = {319--340},
  timestamp = {Mon, 16 Sep 2019 01:00:00 +0200},
  biburl = {https://dblp.org/rec/journals/siamcomp/AttiyaR98.bib},
  bibsource = {dblp computer science bibliography, https://dblp.org},
  _bib2doi_selected = {dblp:/rec/journals/siamcomp/AttiyaR98.bib},
  _bib2doi_confirmed = {true},
}

@book{AttiyaW2004,
  title = {Distributed computing --- fundamentals, simulations, and advanced topics},
  author = {Hagit Attiya and Jennifer L. Welch},
  edition = {2},
  year = {2004},
  publisher = {Wiley},
  timestamp = {Mon, 15 Jul 2019 01:00:00 +0200},
  biburl = {https://dblp.org/rec/books/daglib/0017536.bib},
  bibsource = {dblp computer science bibliography, https://dblp.org},
  isbn = {978-0-471-45324-6},
  series = {Wiley series on parallel and distributed computing},
  _bib2doi_selected = {dblp:/rec/books/daglib/0017536.bib},
  _bib2doi_confirmed = {true},
  _bib2doi_finished = {true},
}

@article{bracha_asynchronous_1987,
  title = {Asynchronous {Byzantine} agreement protocols},
  volume = {75},
  issn = {0890-5401},
  doi = {10.1016/0890-5401(87)90054-X},
  number = {2},
  urldate = {2024-03-08},
  journal = {Information and Computation},
  author = {Bracha, Gabriel},
  month = nov,
  year = {1987},
  pages = {130--143},
  timestamp = {Fri, 12 Feb 2021 00:00:00 +0100},
  biburl = {https://dblp.org/rec/journals/iandc/Bracha87.bib},
  bibsource = {dblp computer science bibliography, https://dblp.org},
  _bib2doi_selected = {dblp:/rec/journals/iandc/Bracha87.bib},
  _bib2doi_confirmed = {true},
}

@inproceedings{civit2022crime,
  title={Crime and punishment in distributed byzantine decision tasks},
  author={Civit, Pierre and Gilbert, Seth and Gramoli, Vincent and Guerraoui, Rachid and Komatovic, Jovan and Milosevic, Zarko and Seredinschi, Adi},
  booktitle={IEEE 42nd International Conference on Distributed Computing Systems (ICDCS)},
  pages={34--44},
  year={2022},
  doi = {10.1109/ICDCS54860.2022.00013}
}

@article{Coan1988,
  author = {Brian A. Coan},
  title = {A Compiler that Increases the Fault Tolerance of Asynchronous Protocols},
  journal = {IEEE Transactions on Computers},
  month = dec,
  year = {1988},
  publisher = {IEEE Computer Society},
  address = {USA},
  volume = {37},
  number = {12},
  issn = {0018-9340},
  url = {https://doi.org/10.1109/12.9732},
  doi = {10.1109/12.9732},
  pages = {1541--1553},
  timestamp = {Mon, 28 Aug 2023 01:00:00 +0200},
  biburl = {https://dblp.org/rec/journals/tc/Coan88.bib},
  bibsource = {dblp computer science bibliography, https://dblp.org},
  _bib2doi_selected = {dblp:/rec/journals/tc/Coan88.bib},
  _bib2doi_confirmed = {true},
  _bib2doi_finished = {true},
}

@article{dwork_consensus_1988,
  title = {Consensus in the presence of partial synchrony},
  volume = {35},
  issn = {0004-5411},
  doi = {10.1145/42282.42283},
  number = {2},
  urldate = {2024-03-27},
  journal = {Journal of the ACM},
  author = {Dwork, Cynthia and Lynch, Nancy and Stockmeyer, Larry},
  month = apr,
  year = {1988},
  pages = {288--323},
  timestamp = {Fri, 24 Mar 2023 00:00:00 +0100},
  biburl = {https://dblp.org/rec/journals/jacm/DworkLS88.bib},
  bibsource = {dblp computer science bibliography, https://dblp.org},
  _bib2doi_selected = {dblp:/rec/journals/jacm/DworkLS88.bib},
  _bib2doi_confirmed = {true},
}

@inproceedings{faleiro_generalized_2012,
  title = {Generalized lattice agreement},
  isbn = {978-1-4503-1450-3},
  doi = {10.1145/2332432.2332458},
  language = {en},
  urldate = {2024-06-07},
  booktitle = { {ACM} 31st symposium on {Principles} of distributed computing (PODC)},
  publisher = {ACM},
  author = {Faleiro, Jose M. and Rajamani, Sriram and Rajan, Kaushik and Ramalingam, G. and Vaswani, Kapil},
  year = {2012},
  pages = {125--134},
  timestamp = {Sun, 06 Oct 2024 01:00:00 +0200},
  biburl = {https://dblp.org/rec/conf/podc/FalerioRRRV12.bib},
  bibsource = {dblp computer science bibliography, https://dblp.org},
  _bib2doi_selected = {dblp:/rec/conf/podc/FalerioRRRV12.bib},
  _bib2doi_confirmed = {true},
}

@inproceedings{di_luna_synchronous_2020,
  author = {Giuseppe Antonio Di Luna and Emmanuelle Anceaume and Silvia Bonomi and Leonardo Querzoni},
  title = {Synchronous {Byzantine} Lattice Agreement in {$O(\log f)$} Rounds},
  booktitle = {IEEE 40th International Conference on Distributed Computing Systems (ICDCS)},
  pages = {146--156},
  publisher = {{IEEE}},
  year = {2020},
  timestamp = {Mon, 03 Jan 2022 00:00:00 +0100},
  biburl = {https://dblp.org/rec/conf/icdcs/LunaABQ20.bib},
  bibsource = {dblp computer science bibliography, https://dblp.org},
  doi = {10.1109/ICDCS47774.2020.00056},
  _bib2doi_selected = {dblp:/rec/conf/icdcs/LunaABQ20.bib},
  _bib2doi_confirmed = {true},
}

@inproceedings{di_luna_byzantine_2020,
  title = {Byzantine generalized lattice agreement},
  author = {Giuseppe Antonio Di Luna and Emmanuelle Anceaume and Leonardo Querzoni},
  booktitle = {IEEE International Parallel and Distributed Processing Symposium (IPDPS)},
  pages = {674--683},
  year = {2020},
 timestamp = {Mon, 03 Jan 2022 00:00:00 +0100},
  biburl = {https://dblp.org/rec/conf/ipps/LunaAQ20.bib},
  bibsource = {dblp computer science bibliography, https://dblp.org},
  doi = {10.1109/IPDPS47924.2020.00075},
  url = {https://doi.org/10.1109/IPDPS47924.2020.00075},
  _bib2doi_selected = {dblp:/rec/conf/ipps/LunaAQ20.bib},
  _bib2doi_confirmed = {true},
  _bib2doi_finished = {true},
}

@inproceedings{ShapiroPBZ2011,
  title = {Conflict-free replicated data types},
  author = {Shapiro, Marc and Pregui{\c{c}}a, Nuno and Baquero, Carlos and Zawirski, Marek},
  booktitle = {Symposium on Self-Stabilizing Systems},
  pages = {386--400},
  year = {2011},
  organization = {Springer},
  timestamp = {Tue, 07 May 2024 01:00:00 +0200},
  biburl = {https://dblp.org/rec/conf/sss/ShapiroPBZ11.bib},
  bibsource = {dblp computer science bibliography, https://dblp.org},
  doi = {10.1007/978-3-642-24550-3_29},
  _bib2doi_selected = {dblp:/rec/conf/sss/ShapiroPBZ11.bib},
  _bib2doi_confirmed = {true},
}

@inproceedings{zheng_byzantine_2020_synchronous,
  title = {Byzantine Lattice Agreement in Synchronous Message Passing Systems},
  author = {Xiong Zheng and Vijay K. Garg},
  booktitle = {34th International Symposium on Distributed Computing (DISC)},
  year = {2020},
  publisher = {Schloss Dagstuhl - Leibniz-Zentrum f{\"{u}}r Informatik},
  volume = {179},
  pages = {32:1--32:16},
  timestamp = {Wed, 07 Oct 2020 01:00:00 +0200},
  biburl = {https://dblp.org/rec/conf/wdag/ZhengG20.bib},
  bibsource = {dblp computer science bibliography, https://dblp.org},
  doi = {10.4230/LIPIcs.DISC.2020.32},
  url = {https://doi.org/10.4230/LIPIcs.DISC.2020.32},
  series = {LIPIcs},
  _bib2doi_selected = {dblp:/rec/conf/wdag/ZhengG20.bib},
  _bib2doi_confirmed = {true},
  _bib2doi_finished = {true},
}

@inproceedings{zheng_byzantine_2020_asynchronous,
  author = {Xiong Zheng and Vijay K. Garg},
  title = {Byzantine Lattice Agreement in Asynchronous Systems},
  booktitle = {24th International Conference on Principles of Distributed Systems (OPODIS)},
  pages = {4:1--4:16},
  series = {LIPIcs},
  isbn = {978-3-95977-176-4},
  issn = {1868-8969},
  year = {2020},
  volume = {184},
  publisher = {Schloss Dagstuhl - Leibniz-Zentrum f{\"{u}}r Informatik},
  url = {https://doi.org/10.4230/LIPIcs.OPODIS.2020.4},
  urn = {urn:nbn:de:0030-drops-134894},
  doi = {10.4230/LIPIcs.OPODIS.2020.4},
  annote = {Keywords: Byzantine Lattice Agreement, Asynchronous},
  timestamp = {Mon, 25 Jan 2021 00:00:00 +0100},
  biburl = {https://dblp.org/rec/conf/opodis/ZhengG20.bib},
  bibsource = {dblp computer science bibliography, https://dblp.org},
  _bib2doi_selected = {dblp:/rec/conf/opodis/ZhengG20.bib},
  _bib2doi_confirmed = {true},
}

@inproceedings{zheng_linearizable_2018,
  author = {Zheng, Xiong and Garg, Vijay K. and Kaippallimalil, John},
  title = {{Linearizable replicated state machines with lattice agreement}},
  booktitle = {23rd International Conference on Principles of Distributed Systems (OPODIS)},
  pages = {29:1--29:16},
  series = {LIPIcs},
  isbn = {978-3-95977-133-7},
  issn = {1868-8969},
  year = {2020},
  volume = {153},
  publisher = {Schloss Dagstuhl -- Leibniz-Zentrum f{\"u}r Informatik},
  urn = {urn:nbn:de:0030-drops-118158},
  doi = {10.4230/LIPIcs.OPODIS.2019.29},
  timestamp = {Wed, 12 Feb 2020 00:00:00 +0100},
  biburl = {https://dblp.org/rec/conf/opodis/ZhengGK19.bib},
  bibsource = {dblp computer science bibliography, https://dblp.org},
  _bib2doi_selected = {dblp:/rec/conf/opodis/ZhengGK19.bib},
  _bib2doi_confirmed = {true},
}
